\documentclass{article} 
\usepackage[preprint]{colm2026_conference}

\usepackage{microtype}
\usepackage{hyperref}
\usepackage{url}
\usepackage{booktabs}

\usepackage{graphicx}
\usepackage{subcaption}
\usepackage{subcaption}
\usepackage{multirow}
\usepackage{wrapfig}
\usepackage[most]{tcolorbox} 
\usepackage{listings}
\usepackage[T1]{fontenc}

\usepackage{lineno}

\definecolor{darkblue}{rgb}{0, 0, 0.5}
\hypersetup{colorlinks=true, citecolor=darkblue, linkcolor=darkblue, urlcolor=darkblue}

\author{
Jinyao Guo$^{1}$, Chengpeng Wang$^{1}$, Dominic Deluca$^{1}$, Jinjie Liu$^{2}$, Zhuo Zhang$^{3}$, Xiangyu Zhang$^{1}$ \\
$^{1}$Purdue University, $^{2}$University of Southern California, $^{3}$Columbia University \\
\texttt{\{guo287, wang2856, ddeluca, xyzhang\}@purdue.edu, jinjiel@usc.edu, zz2736@columbia.edu}
}

\title{\toolname: Learn to Detect Bugs Like Human}

\newcommand{\toolname}{\textsc{BugScope}}
\newcommand{\buggypattern}{anti-pattern}
\newcommand{\Buggypattern}{Anti-pattern}
\newcommand{\buggypatterns}{anti-patterns}

\newcommand{\circlednum}[1]{%
    \tikz[baseline=(char.base)]{
        \node[shape=circle, draw, fill=black, text=white, inner sep=0.6pt, minimum size=1.4ex] (char) {\textbf{#1}};
    }%
}

\newif\ifshowcomments
\showcommentsfalse

\ifshowcomments
\newcommand{\wcp}[1]{\textcolor{blue}{[chengpeng: #1]}}
\newcommand{\gjy}[1]{\textcolor{brown}{[jinyao: #1]}}
\newcommand{\xz}[1]{\textcolor{red}{[Xiangyu: #1]}}
\newcommand{\zz}[1]{\textcolor{pink}{[Zhuo: #1]}}
\else
\newcommand{\wcp}[1]{}
\newcommand{\gjy}[1]{}
\newcommand{\xz}[1]{}
\newcommand{\zz}[1]{}
\fi

\newif\ifrevmode
\revmodetrue

\ifrevmode
  
\else
  
\fi

\begin{document}

\ifcolmsubmission
\linenumbers
\fi

\maketitle

\begin{abstract}
Software auditing is an increasingly critical task in the era of rapid code generation. While LLM-based auditors have demonstrated strong potential, their effectiveness remains limited by misalignment with the highly complex, domain-specific nature of bug detection. 
In this work, we introduce \toolname{}, a framework that mirrors how human auditors learn specific bug patterns from representative examples and apply this knowledge during code auditing.
\toolname{} structures auditing into three steps: seed identification, context retrieval, and bug detection, and aligns LLMs to each step by analyzing real bug reports and mutated examples, and distilling concise, reusable guidelines.
On a curated dataset of 33 real-world bugs from 21 widely used open-source projects, \toolname{} achieves 86.05\% precision and 87.88\% recall, corresponding to an F1 score of 0.87. By comparison, leading industrial tools such as Claude Code (with Claude Opus 4.6) and Cursor BugBot achieve F1 scores of only 0.51 and 0.43, respectively. Beyond benchmarks, large-scale evaluation on real-world projects such as the Linux kernel uncovered 184 previously unknown bugs, of which 78 have already been fixed and 7 explicitly confirmed by developers. Our code is available at 
(\url{https://github.com/jinyaoguo/BugScope}).

\end{abstract}

\section{Introduction}
\label{sec:introduction}

%
%



Bugs pose a serious threat to software systems. With recent advances in GenAI-driven code generation, code is now being produced at an unprecedented pace, which makes code auditing more critical than ever. Traditionally, auditing has relied on static and symbolic program analysis tools such as Infer~\citep{Infer} and CodeQL~\citep{CodeQL}. These tools are effective for detecting certain classes of bugs, but they often fall short when deeper semantic reasoning is required, as they typically abstract programs into graphs and perform purely structural reasoning while ignoring code semantics.
LLM-based auditors, such as BugBot~\citep{cursorbugbot}, show significant promise since large reasoning models can reason about semantic information more deeply. However, they remain constrained by limited context lengths and a tendency to hallucinate in the presence of complex program structures. 

The root cause of the difficulty in using LLMs for code auditing is that they are not inherently aligned with this complex, domain-specific task. The Common Weakness Enumeration (CWE), a widely adopted industry standard, categorizes software weaknesses into over 900 distinct types~\citep{MITRE_CWE}, each corresponding to violations of fundamental software properties. Even within a single category, bugs can arise in diverse semantic contexts and manifest through varied \emph{\buggypatterns}, further complicating detection. In such settings, general-purpose LLM-based auditors often produce excessive false positives or fail to identify true bugs.
In contrast, human auditors learn from historical bug reports, internalizing recurring \buggypatterns{} (both generic and system-specific) through repeated exposure. They then review new code for suspicious patterns with this knowledge, retrieve relevant context, and perform deeper semantic reasoning to verify the presence of bugs.


Inspired by the manual auditing process, we propose a new technique that automatically customizes detection logic for diverse \buggypatterns{}. By integrating structured workflows, domain-specific constraints, and synthesized guidelines, our approach enables the model to reason about code semantics systematically and reliably, improving bug detection effectiveness without extra human effort.
Specifically, we begin by designing a workflow that mirrors how human experts conduct code audits. 
Instead of attempting to reason about a codebase holistically, auditors follow a disciplined workflow of seed identification, context retrieval, and bug detection.

In the {\em learning stage}, we provide the model with representative bug cases and guide it to generate additional and complementary examples along each step of this workflow. 
For seed identification, the model enumerates typical program constructs that instantiate potential bug locations, e.g., division operator for divide-by-zero (DBZ) bugs. 
For the retrieval step, the model explores the different ways of gathering relevant code, such as following {\em data dependences} to variable definitions, or searching usage sites of specific APIs. To make retrieval cost-effective, we introduce {\em termination conditions}, which specify when further expansion of the code context is unnecessary. 
In the detection step, the model explores {\em evidence patterns} for bugs, concise code patterns that enable an auditor to conclude a bug with high confidence. For example, evidence for DBZ includes the explicit assignment of zero to a variable or the loading of an external value into a variable, where the latter may evaluate to zero depending on the input. By codifying such evidence, we align the model’s focus and reasoning discipline with the requirements of real-world bug detection.

From the generated examples, \toolname{} synthesizes two sets of guidelines: (1) a {\em retrieval strategy}, consisting of a seed extractor and traversal rules for a {\em context retrieval agent}, and (2) a {\em detection prompt}, consisting of anti-pattern specifications for a {\em bug detection agent}. In the subsequent {\em auditing stage}, these agents collaborate to analyze new projects: the retrieval agent identifies seeds and constructs candidate contexts using the retrieval strategy, while the detection agent applies the detection prompt to those contexts to generate bug reports.

\begin{figure*}[t]
	\centering
	\includegraphics[width=0.98\linewidth]{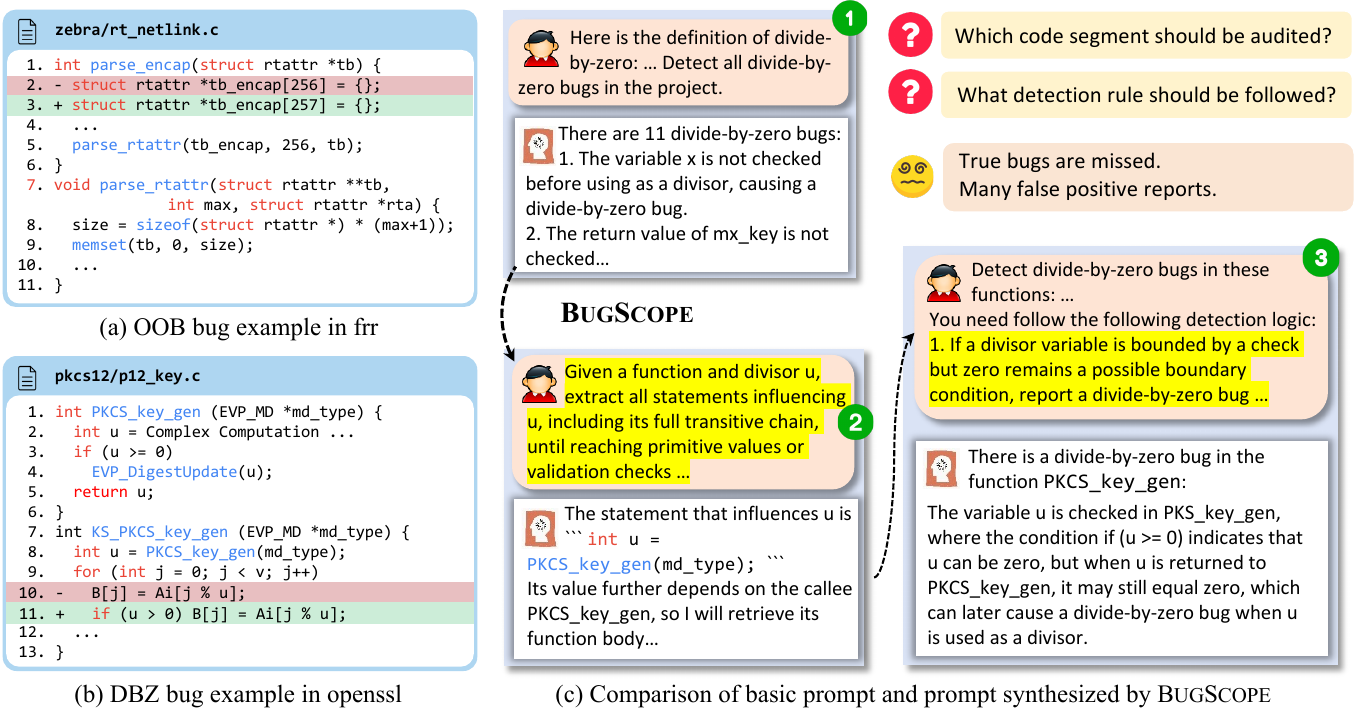}
	\vspace{-3mm}
	\caption{Motivating examples}
    \vspace{-5mm}
	\label{fig:examplecode}
\end{figure*}

\noindent{\bf Results.} We evaluate \toolname{} using three reasoning models, namely Claude 3.7 Sonnet Thinking, OpenAI o4-mini, and DeepSeek-R1. Experimental results demonstrate that \toolname{} consistently outperforms other LLM-driven code auditors like RepoAudit and Claude Code, as well as commercial static analysis tools, including Meta Infer, Cursor Bugbot, and CodeRabbit. When powered by Claude 3.7 Sonnet Thinking, \toolname{} achieves 86.05\% precision, 87.88\% recall, and an F1 score of 0.87 in a controlled experiment. In contrast, the highest F1 score among the baselines is only 0.51. 
For specific \buggypatterns{}, such as negative offset and insufficient zero check, the baselines detect at most one true positive and several of them even fail to identify any buggy cases.
When deployed on real-world codebases, \toolname{} uncovers 184 new bugs, 78 of which have already been fixed and 7 confirmed by developers. Notably, twelve bugs in the Linux kernel discovered by \toolname{} have been either fixed or acknowledged by maintainers. These results highlight the practical effectiveness and applicability of \toolname{} in real-world software auditing. 


\section{Motivation}
\vspace{-1mm}




Bug detection is an extremely complex domain-specific task. In this section, we use two real examples to illustrate such complexity and the entailed challenges for LLMs and agents. The code snippets have been substantially simplified for illustration. 
Figure~\ref{fig:examplecode}(a) illustrates an out-of-bound (OOB) bug. The original buggy code at line 2 declares a local buffer of size 256, which is passed to function \verb|parse_rtattr|() at line 5.
However, inside the function (line 10), \verb|memset|() is invoked with the size of 257. The two functions involved reside in different files. Figure~\ref{fig:examplecode}(b) presents a divide-by-zero (DBZ) bug. The variable \verb|u| is obtained through complex computation and may yield a zero value. It is later used as a divisor at line 10.


\noindent
{\bf Limitations of Existing Techniques}
Although these two cases are of typical bug types: OOB and DBZ, traditional static analysis tools~\citep{Xue16SVF, Arzt14FlowDroid, CodeQL, Infer} have difficulty detecting them due to tools' inherent limitations in reasoning about deep semantics. In particular, these tools usually ignore semantics in symbols, model programs as graphs, and leverage iterative graph algorithms to derive analysis results. As such, tool developers have to explicitly model a lot of semantics (e.g., manually constructing symbolic summaries of library functions), which may not be comprehensive to cover a wide spectrum of projects. For example, without the code of \verb|memset|(), which is the typical case as it belongs to a library, or the manually created model for the function, classic tools cannot determine an OOB bug in the first case. Without the ability to reason about the mathematical semantics of the computation of \verb|u| in the second case, the DBZ bug is missed as there is not a zero value assignment of \verb|u|.

Second, although LLMs have demonstrated the ability to reason about deep semantics (e.g., in pointer aliasing~\citep{guo2025repoaudit}), they fall short in detecting these bugs. As shown in Figure~\ref{fig:examplecode}(c) step \circlednum{1}, simply providing the bug definition and asking the LLM (Claude 3.7 in our case) to inspect the given code snippet, like in existing code audit agents such as Cursor Bugbot and CodeRabbit, can hardly work. It faces multiple challenges: {\em (1) retrieving sufficient code fragments in the project for bug detection, and (2) determining true bugs from numerous candidates}. As shown in the conversation box \circlednum{1}, the LLM reports 11 DBZ for the code snippet in (b) with 10 false positives as it considers all divisions as potentially buggy when it cannot find a zero-value check in the surrounding code region. If we change the prompt to forbid the LLM to report any DBZ bugs unless it has explicit evidence that the divisor may be zero,  it reports no bug. LLMs have similar problems with the OOB example. It either reports a lot of false positives (e.g., regarding pointer \verb|rta| at line 8) or misses the real bug. Particularly, it is extremely difficult for the LLM or agent to retrieve the exactly needed contexts for the numerous pointers and buffers in the code snippet for detection. 



Third, recent research explores integrating LLMs with traditional static analysis~\citep{wangllmdfa, guo2025repoaudit, naik2021sporq, li2025automatedstaticvulnerabilitydetection, yang2025knightertransformingstaticanalysis} to mitigate the above problems, e.g., by using low-level data-flow to retrieve relevant code and enforcing symbolic rules for bug detection. However, such retrieval often has a fixed bound (such as three layers of function calls) to control cost and the rules are often too restrictive due to their symbolic nature. For example, RepoAudit uses a retrieval bound of 3 and only considers DBZ that initiates from a 0 value assignment. As a result, it misses the two example bugs. 


\begin{tcolorbox}[colback=gray!10, colframe=black,
  boxrule=0.3pt, arc=2mm, left=1mm, right=1mm, top=0.5mm, bottom=0.5mm,
  enhanced
  ]
\noindent

The core limitation is that LLMs are not inherently aligned with complex, domain-specific tasks such as bug detection. Without clear detection logic, the model may become overly restrictive, reducing utility by missing real bugs, or overly permissive, diminishing effectiveness by generating excessive false positives.\end{tcolorbox}

\vspace{-1mm}



\smallskip
\noindent
{\bf Our Idea.}

Inspired by the manual detection workflow, \toolname{} adopts the same discipline used by human auditors: reasoning through a step-by-step process and progressively gathering only the code relevant to a suspected issue.
Unlike the holistic approach adopted by most existing LLM-based review agents (e.g., Bugbot, CodeRabbit), where the model is given an entire code segment and asked to uncover all possible bugs, our workflow narrows the model’s attention to a focused set of closely related bug instances (e.g., all memory-safety issues involving a particular buffer). This targeted strategy both reduces hallucinations and improves reliability.

In addition to simplifying the workflow to ease detection, we present challenging examples, i.e., real bug cases with complex causal chains, and prompt a strong reasoning model to reflect on them. The model engages in deeper analysis and generalization to uncover relevant prior knowledge (e.g., mutate code patterns that imply similar bugs) and then distills guidelines for each stage of the workflow. This mirrors how human experts calibrate their alertness and rigor by studying historical bug reports before conducting real audits.

Considering DBZ bugs, \toolname{} synthesizes a number of guidelines from the example. The highlighted texts in conversations \circlednum{2} and \circlednum{3} show the retrieval and detection guidelines. 
Particularly, the retrieval guideline indicates that the agent needs to bring in all the code covering the entire transitive computation of \verb|u|'s value. 
The bug detection guideline indicates that if 0 is a boundary condition for a variable, the variable may hold a zero value although there is no explicit zero value assignment to the variable, which resembles how a human expert leverages indirect hints during auditing. 
These guidelines are precise and effective. They are used to detect DBZ in 12 {\em other} projects and report 23 true bugs with only 5 false positives. Similar results for other bug types are reported in Section~\ref{sec:reproduce}. 
We will explain more details in the following sections.

\section{Approach}
\label{sec:technical}

\begin{figure*}[t]
    \centering
    \includegraphics[width=0.95\linewidth]{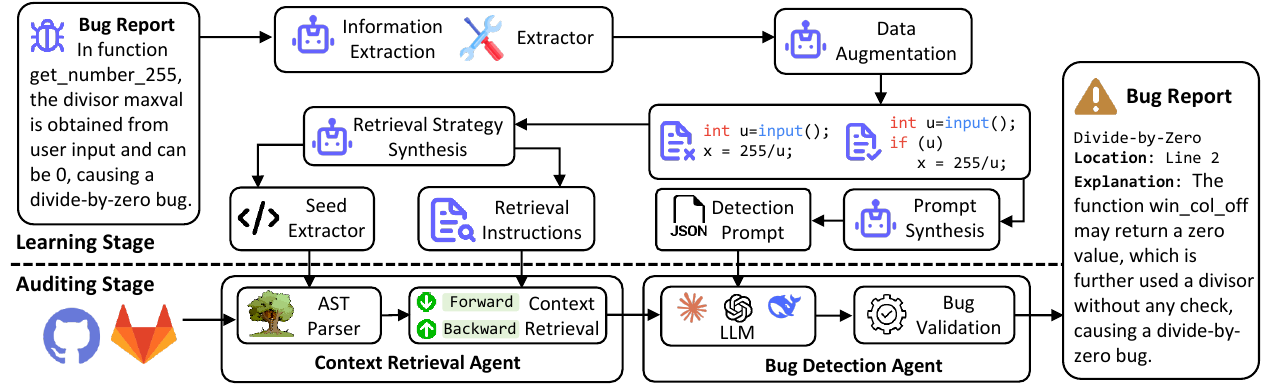}
    \caption{The workflow of \toolname{}
    }
    \vspace{-5mm}
    \label{fig:workflow}
\end{figure*}

Figure~\ref{fig:workflow} illustrates the overview of \toolname{}. It consists of two phases as separated by the dashed line: a {\em learning phase} that automatically derives a set of guidelines for code retrieval and bug detection from a given bug report and an {\em auditing stage} in which the guidelines are used to instruct two interacting agents for bug detection. 

In the learning stage, \toolname{} takes an existing bug report as input, extracts descriptive details and relevant code fragments, and guides the LLM to reason through the structured three-step workflow mentioned in the last section.
The LLM is further prompted to produce both positive and negative examples illustrating complementary cases across these steps. From these examples, the system abstracts (1) a \emph{retrieval strategy}, consisting of a seed extractor and retrieval rules for the {\it context retrieval agent}, and (2) a \emph{detection prompt}, consisting of detection guidelines for the {\it bug detection agent}.
In the auditing stage, given a repository, the retrieval agent identifies potential seeds and gathers associated code fragments to construct candidate contexts in line with the learned guidelines. The detection agent then applies the detection prompt to the contexts to detect bugs. The following subsections describe the design of each stage in detail.

\subsection{Learning Stage}
\label{sec:learning}
The goal of the learning stage is to derive the principles/guidelines for a particular kind of bugs. It comprises four steps: {\em information extraction}, {\em data augmentation}, {\em seed extraction and retrieval guideline synthesis} and {\em detection guideline synthesis}.

\begin{figure*}[t]
    \centering
    \vspace{-1mm}
    \includegraphics[width=0.92\linewidth]{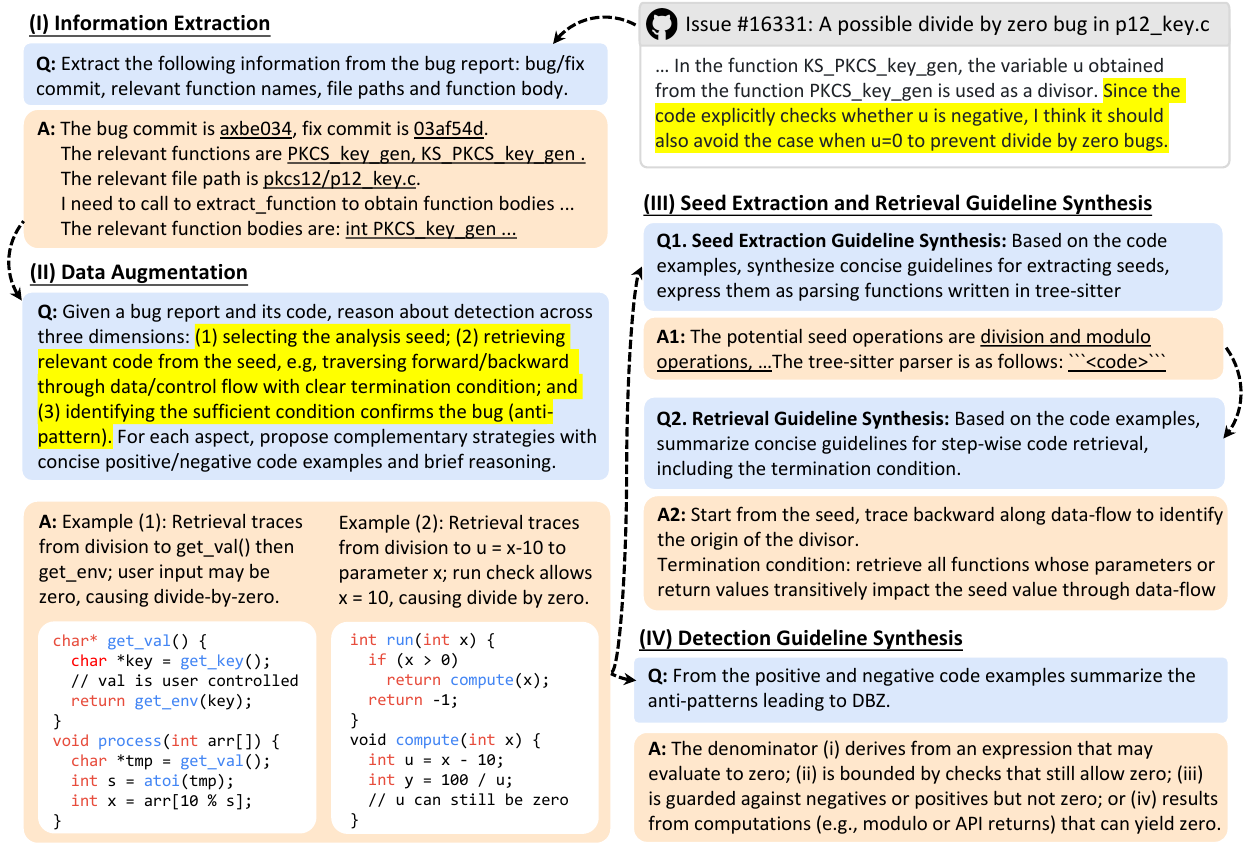}
    \caption{An example of guideline derivation
    }
    \vspace{-5mm}
    \label{fig:example}
\end{figure*}

\noindent
\textbf{Information Extraction.}
\toolname{} initiates guideline derivation from an existing bug report of a specific \buggypattern{}.
Since bug reports are often written in arbitrary formats and may lack critical details (e.g., relevant code snippets or explicit reasoning steps), we first analyze the bug pattern and extract the necessary information from the associated code repository for the following steps. As shown in Step (I) of Figure~\ref{fig:example}, an information-extraction agent processes the bug report and retrieves fix commits, relevant function names, file paths, and pre-/post-fix function bodies through a tool-assisted code lookup mechanism. The detailed implementation descriptions are provided in Appendix~\ref{appendix:info-extraction}.

\noindent
{\bf Data Augmentation.}
\toolname{} begins with a single example for each bug type. While such an example captures some complexity, it is insufficient to represent the full spectrum of possible manifestations. Directly prompting the model to derive detection guidance from a single example often performs poorly: the resulting guidance is either too specific to generalize across variations of the same \buggypattern{}, or too general, leading to excessive false positives.
To derive precise and effective guidance, we decompose bug detection into the three stages of our workflow and instruct the model to explore variations along each stage. The corresponding prompt is shown in the ``Data Augmentation'' conversation in Figure~\ref{fig:example}, where the three stages are highlighted in yellow.

\underline{First}, the model is prompted to enumerate {\it seeds} for an \buggypattern{} and generate illustrative examples. For instance, seeds for DBZ bugs include division and modulo operations, while seeds for OOB bugs include buffer indexing, pointer dereferencing after pointer arithmetic and buffer access APIs. Without considering further complexities, the model can enumerate most cases along this dimension. Examples (1) and (2) in Figure~\ref{fig:example} (white text boxes) illustrate two seeds for DBZ bugs.

\underline{Second}, the model is instructed to consider how to retrieve relevant code snippets starting from a seed and when to terminate such retrieval. This process is bug-type specific. For DBZ bugs, relevant code should be collected by tracing {\em data dependence} and {\em control dependence} relations in a {\em backward} fashion. That is, beginning with the seed, one should retrieve variable definitions that contribute, directly or indirectly, to the seed through computations or guarding conditions. In Example (2) of Figure~\ref{fig:example}, starting from the division on \verb|u|, the retrieval should transitively include the computation of \verb|u| in the caller function \verb|run|(), especially the guarding condition \verb|x > 0|, which determines whether the division executes. Other retrieval strategies include following {\em control flow} in a {\em forward} fashion, as in memory leak (MLK) detection: beginning from an allocation point, one follows program paths forward to check whether the allocated resource is eventually released.

The {\it termination condition} is essential for cost-effective auditing and varies by bug type too. For DBZ retrieval, a naive strategy is to follow all transitive data dependences until the variable initializations. However, this is often inefficient and unnecessary. In practice, human auditors rely on domain knowledge to terminate earlier. For example, in Example (1) of Figure~\ref{fig:example}, starting from the seed \verb|10 % s|, the retrieval proceeds to the callee \verb|get_val| because \verb|s| is transitively derived from its return value \verb|tmp|. Inside the callee, the return value is obtained through a library call that reads an environment variable using \verb|key|. At this point, retrieving the computation of \verb|key| is unnecessary, since the auditor can conclude that \verb|val| may be zero due to the external call.

\underline{Third}, the model is instructed to generate examples of possible {\em sufficient conditions} for a given bug type, namely code patterns that enable an auditor to draw a highly confident conclusion about the presence of a bug. We refer to these patterns as {\em evidence patterns} in this paper. In practice, when auditing code for a particular class of bugs, human experts rarely rely solely on formal definitions; instead, they actively seek such evidence patterns. For example, an overly restrictive rule used by RepoAudit for DBZ detection involved traversing backward from a division operation and checking for a direct zero assignment, thereby avoiding the complexity of reasoning about variable value ranges. 

In reality, auditors consider a broader set of anti-patterns. One example is the boundary check \verb|u >= 0| in Figure~\ref{fig:examplecode}(b), which implicitly reveals that 0 is treated as a boundary value, as noted in the corresponding issue report. Another instance, shown in Example (2) of Figure~\ref{fig:example}, is the guard \verb|x > 0|, which suggests that \verb|x| is positive and may later produce a zero value when evaluated as \verb|x - 10|. 
A further anti-pattern for OOB appears in \verb|calloc(n, sizeof(T))|, where the computed size \verb|n * sizeof(T)| may overflow and wrap to a smaller value, causing a buffer overflow.

In addition to generating buggy (positive) variances, the LLM is also directed to produce similar but correct (negative) examples, enabling contrastive learning. Each example is paired with a brief explanation. Compared to relying solely on real issue reports, our data augmentation achieves greater diversity by explicitly guiding the model to sample along the three dimensions. Moreover, the generated code examples tend to be concise and focused, avoiding much of the noise present in real-world code, which in turn facilitates more effective downstream guideline synthesis.

\noindent
{\bf Guideline Synthesis.}
In this step, we leverage existing bug reports and augmented examples to synthesize guidelines for the three-step bug detection workflow, including a seed extractor, a set of retrieval guidelines, and detection guidelines.

The seed extractor is synthesized based on the parsing library Tree-sitter~\citep{brunsfeld2018tree} to identify seed locations. For example, a seed extractor for DBZ bugs can be implemented using a Tree-sitter query that locates division or modulo expressions, e.g., \verb|binary_expression [operator: "/" or "%"]|. The synthesized seed extractor is then validated and refined on the examples to ensure accurate identification of target operations.

In addition, a set of retrieval guidelines is synthesized to specify how to collect program statements relevant to the identified seeds and when to terminate the retrieval process. These guidelines capture three categories of information: (1) relations to follow during retrieval, such as function invocations, data dependencies, and control dependencies; (2) the direction of traversal, including forward, backward, or bidirectional; and (3) stopping conditions. A representative guideline is illustrated in Step~(III) of Figure~\ref{fig:example}.
Stopping conditions vary across bug types. For DBZ, one synthesized guideline is to terminate at an anti-pattern, as further retrieval becomes unnecessary once sufficient evidence is identified. For memory safety bugs such as out-of-bounds (OOB), a typical stopping condition is to stop at the enclosing function of the buffer variable under inspection. For example, consider a local buffer defined in function \verb|A()| that is passed to \verb|B()| and subsequently to \verb|C()|, where the buffer is accessed. The guideline prescribes backward traversal starting from the buffer accesses (i.e., the seeds) along the call chain until reaching the enclosing function \verb|A()|. Together with its callees, \verb|A()| provides sufficient context for OOB detection.

The synthesized bug detection guidelines describe the evidence patterns required to report a bug. Step~(IV) in Figure~\ref{fig:example} illustrates the synthesis process with examples for DBZ bug detection. To ensure that the guidance is precise and effective, the synthesized guidelines are validated on the augmented examples, where misclassified cases are analyzed by the synthesizer and used to iteratively refine the guidelines until all cases are correctly detected.

\vspace{-1mm}
\subsection{Auditing Stage}
\label{sec:auditing}
In the auditing stage, \toolname{} applies a multi-agent framework to perform precise context retrieval and bug detection follows the guidelines synthesized before. 
Given a project and a target bug type, \toolname{} first invokes the {\it context retrieval agent} to identify code fragments that are likely to contain bugs. These candidate fragments are then passed to the {\it bug detection agent}, which generates bug reports.

The goal of the context-retrieval agent is to scan the codebase, identify suspicious operations, and collect sufficient calling context for precise bug detection. However, real-world projects are large in scale (with an average of 81K lines of code in our collected dataset), making it infeasible for LLMs to analyze the entire codebase directly.
To address this challenge, the context-retrieval agent follows a neural-symbolic procedure that combines static program structure analysis with LLM-based semantic reasoning. Specifically, we first parse the repository using Tree-sitter~\citep{brunsfeld2018tree} to construct the abstract syntax tree (AST) and an inter-procedural call graph. The synthesized seed extractor is then applied to identify suspicious operations from the constructed AST. Starting from each seed statement, the agent incrementally extracts relevant statements as detection context according to the context retrieval guidelines.

To handle the explosion problem of inter-procedural dependencies across the repository, the agent performs selective inter-procedural expansion. Starting from the function containing the seed operation, it first extracts variables that propagate inter-procedural dependencies from the AST, including parameters, arguments, and return values. It then prompts the LLM to analyze the intra-procedural dependencies of the seed operation and determine which inter-procedural dependencies should be tracked. Only functions that are semantically relevant to the seed operation are retrieved. This iterative expansion terminates when further traversal is unnecessary for the property being checked, as determined by the synthesized termination conditions.


Finally, the retrieved context is embedded into a bug detection prompt that incorporates the synthesized detection guidelines, enabling the LLM to reason over the context step by step and identify the evidence patterns required for bug reporting.
When tasked with bug detection, LLMs may hallucinate by introducing assumptions that are not grounded in the code semantics (e.g., assuming that memory will not be released if a function throws an exception). To mitigate such hallucinations, we introduce an additional validator that analyzes each generated bug report, checks its feasibility, and verifies whether all underlying assumptions are justified by the code. A bug report is considered valid only when every assumption is supported by the program semantics, thereby improving detection accuracy.

\vspace{-1mm}
\section{Evaluation}
\label{sec:evaluation}
\vspace{-1mm}

We use the Tree-sitter parsing library~\citep{brunsfeld2018tree} to support the context retrieval agent, enabling accurate and scalable syntactic analysis. 
We evaluate \toolname{} on Claude 3.7 Sonnet Thinking (hereafter Claude 3.7), with a 4,096-token output limit and a 2,048-token reasoning limit.
For baselines, we consider the state-of-the-art LLM-driven bug detection tool RepoAudit~\cite{guo2025repoaudit}, as well as three commercial tools, namely BugBot~\cite{cursorbugbot}, ~\citet{coderabbit}, and Infer~\cite{Infer}.

\subsection{Performance of \toolname{} in 6 Real-world Projects}
To evaluate the effectiveness of \toolname{}, we first align it using existing issues, then applying it to real-world C/C++ projects to detect new bugs. 
We focus on three types of bugs, namely Out-of-Bounds (OOB), Divide-by-Zero (DBZ), and Memory Leak (MLK), which span a wide spectrum of program properties, including pointer-related, numeric, and combined properties, and require distinct detection logic.
For each bug type, we select 2–3 previous issues for the learning phase and merge the guidelines derived from these cases. In total, 7 bugs are used for learning. 
We then apply the synthesized retrieval strategies and detection prompts to 6 high-profile real-world C/C++ projects (up to 1M LoC, avg. 654K) to detect new bugs.
As shown in Table~\ref{table:new_bug}, \toolname{} detects 8 DBZ, 8 OOB, and 64 MLK bugs with 76.92\% precision, surpassing all baselines. Of these bugs, 73 have been confirmed or fixed by developers. More details are provided in Appendix~\ref{appendix:real-world}.

\begin{table*}[htbp]
    \centering
    \small
    \caption{The statistics of \toolname\ upon six real-world projects. 
    \textbf{\#TP} and \textbf{\#FP} indicates the numbers of true positives and false positives, respectively.
    \textbf{Co} and \textbf{Fx} denote the numbers of bugs that are confirmed and fixed, respectively. \textbf{P(\%)} denotes precision.}
    \label{table:new_bug}
    \resizebox{0.98\textwidth}{!}{%
    \begin{tabular}{l|ccccc|ccccc|ccccc}
    \toprule
    \multirow{2}{*}{\textbf{Project}}
        & \multicolumn{5}{c|}{\textbf{OOB}}
        & \multicolumn{5}{c|}{\textbf{DBZ}}
        & \multicolumn{5}{c}{\textbf{MLK}} \\
    \cmidrule(lr){2-6} \cmidrule(lr){7-11} \cmidrule(lr){12-16}
    & \textbf{\#TP} & \textbf{\#FP} & \textbf{Co} & \textbf{Fx} & \textbf{P (\%)}
    & \textbf{\#TP} & \textbf{\#FP} & \textbf{Co} & \textbf{Fx} & \textbf{P (\%)}
    & \textbf{\#TP} & \textbf{\#FP} & \textbf{Co} & \textbf{Fx} & \textbf{P (\%)} \\
    \midrule
    vim        & 4 & 2 & 0 & 4 & 66.67 & 0 & 1 & 0 & 0 & 0.00     & 2  & 1 & 0 & 2 & 66.67 \\
    systemd    & 0 & 0 & 0 & 0 & 0.00     & 2 & 1 & 0 & 2 & 66.67 & 0  & 2 & 0 & 0 & 0.00     \\
    dynamips   & 2 & 0 & 0 & 2 & 100.00   & 0 & 0 & 0 & 0 & 0.00     & 21 & 2 & 0 & 19 & 91.30 \\
    zstd       & 1 & 1 & 0 & 0 & 50.00    & 4 & 1 & 0 & 1 & 80.00    & 18 & 4 & 0 & 18 & 81.82 \\
    openldap   & 0 & 2 & 0 & 0 & 0.00     & 0 & 1 & 0 & 0 & 0.00     & 18 & 5 & 0 & 17 & 78.26 \\
    git        & 1 & 0 & 1 & 0 & 100.00   & 2 & 0 & 2 & 0 & 100.00   & 5  & 1 & 2 & 3 & 83.33 \\
    \midrule
    \textbf{Total} 
               & 8 & 5 & 1 & 6 & 61.54 
               & 8 & 4 & 2 & 3 & 66.67 
               & 64 & 15 & 2 & 59 & 81.01 \\
    \bottomrule
    \end{tabular}%
    }
\vspace{-3mm}
\end{table*}

\subsection{Controlled Experiment on Crafted Dataset}
\label{sec:reproduce}

\paragraph{Dataset.} 
To compare the bug-detection capability of \toolname{} and baselines, we first survey recent works published in top venues in computer security and software engineering, and manually collect the bug reports released by the authors as our dataset~\citep{guo2022precise, guo2024precise, huang2024raisin, guo2025repoaudit, shi2021path, shi2018pinpoint}. This dataset contains 33 cases covering three bug types (OOB, DBZ, and MLK) and seven \buggypatterns{}. Additional details are provided in Appendix~\ref{appendix:dataset}.

\begin{table*}[htbp]
\centering
\caption{Statistics of \toolname{} with Claude 3.7. 
\textbf{AP}: Anti-Pattern (Appendix~\ref{appendix:real-world}).
\textbf{\#C}: number of cases. 
\textbf{\#Seed}: number of extracted seeds. 
\textbf{\#R}: reproduced bugs. 
\textbf{\#N}: new bugs. 
\textbf{P(\%)}, \textbf{R(\%)}, \textbf{F1}: precision, recall, F1 score.}
\label{tab:reproduce}
\resizebox{0.8\textwidth}{!}{
\begin{tabular}{c|c|ccc|cccc|ccc|cc}
\toprule
\textbf{Type} & \textbf{AP} & \textbf{\#C} & \textbf{\#Seed} & \textbf{\#R} & \textbf{\#N} & \textbf{\#TP} & \textbf{\#FP} & \textbf{P(\%)} & \textbf{R(\%)} & \textbf{F1} & \textbf{Time (s)} & \textbf{Cost (\$)} \\
\midrule
\multirow{3}{*}{OOB} & OSO & 4 & 269 & 3 & 1 & 4 & 2 & 66.67 & 75.00 & 0.71 & 5,547 & 32.73 \\
 & NOF & 4 & 161 & 4 & 0 & 4 & 2 & 66.67 & 100.00 & 0.80 & 2,567 & 12.35 \\
 & ASO & 4 & 27  & 4 & 0 & 4 & 0 & 100.00 & 100.00 & 1.00 & 1,700 & 4.78 \\
\cmidrule{1-2}
\multirow{2}{*}{DBZ} & IZC & 8 & 32  & 5 & 1 & 6 & 1 & 85.71 & 62.50 & 0.72 & 1,548 & 4.24 \\
 & LZD & 4 & 16  & 4 & 5 & 9 & 0 & 100.00 & 100.00 & 1.00 & 653   & 1.19 \\
\cmidrule{1-2}
\multirow{2}{*}{MLK} & UEC & 4 & 31  & 4 & 1 & 5 & 1 & 83.33 & 100.00 & 0.91 & 1,523 & 5.83 \\
 & MSC & 5 & 10  & 5 & 0 & 5 & 0 & 100.00 & 100.00 & 1.00 & 1,910 & 2.24 \\
\midrule
\multicolumn{2}{c|}{Total} & 33 & 546 & 29 & 8 & 37 & 6 & \textbf{86.05} & \textbf{87.88} & \textbf{0.87} & 15,448 & 63.36 \\
\bottomrule
\end{tabular}
}
\vspace{-2mm}
\end{table*}

\paragraph{Performance of \toolname{}.}
We evaluate \toolname{} on the above dataset.
To control the cost of evaluation, we limit the seed extraction to the source files containing the original reported bugs, while code from the entire projects may be retrieved.  
We measure precision, recall, and F1 score, as well as the time and financial cost of each model.
As shown in Table~\ref{tab:reproduce}, \toolname{} powered by Claude 3.7 reproduces 29 out of 33 bugs, discovers 8 additional ones, and achieves 86.05\% precision, 87.88\% recall, and an overall F1 score of 0.87. In terms of efficiency, the average detection cost per seed is \$0.12. Results with additional models, including DeepSeek-R1 and OpenAI o4-mini, are reported in Appendix~\ref{appendix:moreLLMs}, where we compare accuracy and cost trade-offs across models.

\paragraph{Comparison with Baselines} 
We compare \toolname{} against Claude Code~\citep{claude_code}, RepoAudit~\citep{guo2025repoaudit}, BugBot~\citep{cursorbugbot}, \citet{coderabbit}, and Infer~\citep{Infer}, using the same dataset and evaluation metrics. We also provide the same learning-stage examples to Claude Code to synthesize guidelines before detection.
As summarized in Table~\ref{tab:baseline}, \toolname{} consistently outperforms all baselines across all categories. Powered by Claude 3.7, \toolname{} achieves an F1 score of 0.87. In contrast, Claude Code achieves only 65.38\% precision, 42.42\% recall and an F1 score of 0.51. RepoAudit reaches 34.88\% precision, 42.42\% recall, and an F1 score of 0.38. BugBot and CodeRabbit yield comparable precision but suffer from poor recall, resulting in F1 scores of 0.43 and 0.29, respectively. Infer detects only a single true bug with an F1 score of 0.05.  

These results highlight the limitations of existing tools: the guidelines directly synthesized by Claude Code tend to overfit to provided code patterns and lack a structured reasoning framework, RepoAudit is constrained by its data-flow paradigm and restrictive rules, BugBot and CodeRabbit lack sufficient context reasoning, and Infer is bound by its fixed checkers and build dependencies. In contrast, \toolname{} generalizes across diverse \buggypatterns{}, delivering both high precision and recall. Detailed per-pattern comparisons are presented in Appendix~\ref{appendix:baseline}, and additional case studies are included in Appendix~\ref{appendix:case_study}.

\paragraph{Ablation Studies.} 
We evaluate two ablations of \toolname{}, namely \emph{NoRetrieval} (removing the context retrieval agent) and \emph{NoAlign} (synthesizing prompts directly from bug reports without the data augmentation and abstraction).  
As shown in Table~\ref{tab:ablation}, removing context retrieval reduces recall to 42.42\% with only 3 new bugs discovered, while removing alignment decreases true positives to 24 and increases false positives to 38, more than six times that of \toolname{}.  
Further analysis and illustrative examples are provided in Appendix~\ref{appendix:ablation}.

\subsection{More bug type.}
To evaluate BugScope’s performance across a broader range of bug types, we align \toolname{} on all seven Linux kernel issues used in prior work~\citep{chen2025seal}, each representing a system-specific \buggypattern{}, and use the aligned agents to guide new retrieval strategies and detection prompts.
On Linux, \toolname{} generalizes system-specific patterns and finds 102 bugs with 91.07\% precision. More details are provided in Appendix~\ref{appendix:linux}.
\section{Related Work}
\label{sec:related}

\noindent
{\bf LLM-driven Bug Detection.}  
Recent studies leverage LLMs for bug detection through two paradigms: enhancing traditional analyzers or building autonomous agents. In the former, LLMs are used to generate queries or specifications that augment static analysis pipelines~\citep{li2025iris, li2024enhancing}. In the latter, LLMs act as the primary detection engine, employing techniques such as iterative prompting, subtask decomposition, or cross-referencing against formal specifications~\citep{guo2025repoaudit, wangllmdfa, zheng2025llmagentfunctionalbug}. While effective, these approaches often target narrow bug classes or require substantial domain expertise. In contrast, our work follows a \emph{learn-from-example} paradigm, enabling LLMs to infer \buggypatterns{} from code examples and achieve broad coverage without specialized analysis.  

\noindent
{\bf Customizable Static Analysis.}  
Static analyzers such as FlowDroid~\citep{Arzt14FlowDroid}, SVF~\citep{Xue16SVF}, Infer~\citep{CalcagnoDOY09}, and KLEE~\citep{CadarDE08} are typically tailored to specific bug classes, and adapting them requires significant effort. CodeQL~\citep{CodeQL} supports customization through user-defined queries but poses a steep learning curve. Recent work has explored automatic checker synthesis from templates or examples~\citep{DBLP:journals/corr/abs-2504-16057, DBLP:journals/corr/abs-2503-09002, NaikMSWNR21}, yet these methods remain tied to symbolic analysis frameworks. Our approach differs by leveraging LLMs to synthesize detection prompts that mimic human reasoning, offering a more flexible and broadly applicable paradigm for bug detection.

\section{Conclusion}
\label{sec:conclusion}This paper introduces \toolname{}, an LLM-based multi-agent framework that emulates the human auditing process to learn from examples and detect software bugs. The approach significantly outperforms existing methods and demonstrates strong effectiveness in real-world evaluations.

\section{Ethics Statement}
\label{sec:ethics}


From an ethical perspective, all issues detected during evaluation are manually validated by the authors before any external communication. Each reported issue is independently reviewed by two authors, with disagreements adjudicated by a third author. Evaluation metrics are computed based on the outcomes of this internal review process: cases deemed false positives by the authors are not reported to developers, but are still included in the false positive counts.
Only cases for which all three authors reach consensus are labeled as true bugs and subsequently reported. Importantly, we do not rely on open-source developers to determine the correctness of detected issues. Developers are contacted only after full internal validation, and no part of our evaluation depends on developer feedback. This design avoids ethical concerns associated with outsourcing validation to open-source maintainers or involving them in unconsented human-subject roles.

\bibliography{Paper/bibfile}
\bibliographystyle{colm2026_conference}
\appendix
\onecolumn

\definecolor{codegreen}{rgb}{0,0.6,0}
\definecolor{codegray}{rgb}{0.5,0.5,0.5}
\definecolor{codepurple}{rgb}{0.58,0,0.82}
\definecolor{backcolour}{rgb}{0.95,0.95,0.92}

\lstdefinelanguage{Prompt}{
	keywords=[1]{}, 
	keywordstyle=[1]\color{black}\bfseries,
	identifierstyle=\color{black},
	sensitive=true,
	commentstyle=\color{black}\ttfamily,
	stringstyle=\color{black}\ttfamily,
	morestring=[b]',
	morestring=[b]"
}
\lstset{
	language=Prompt,
	backgroundcolor=\color{backcolour},
	extendedchars=true,
	basicstyle=\scriptsize\ttfamily,
	showstringspaces=false,
	showspaces=false,
    commentstyle=none,
	numbers=none,
	numberstyle=none,
	numbersep=4pt,
	xleftmargin=2em,
	xrightmargin=1em,
	frame=single,
	framexleftmargin=1em,
	framexrightmargin=1em,
	tabsize=2,
	breaklines=true,
	showtabs=false,
	captionpos=t,
	columns=flexible,
	escapeinside={/*}{*/},          
}

\section{Information Extraction Details}
\label{appendix:info-extraction}

For each issue used in the learning phase, we extract structured information from both the bug report and the corresponding code repository.

\paragraph{Textual Information.}
We first collect the basic textual components of the issue, including the title, description, developer comments, the bug commit, and the associated fix commit. These elements provide essential high-level context and often contain implicit clues regarding the root cause or affected components. For the bug commit, we first attempt to extract the commit identifier directly from the description; if this information is missing, we approximate it by selecting the commit immediately preceding the issue creation time. This heuristic maximizes the likelihood that the extracted code corresponds to the version referenced in the bug report.

\paragraph{Semantic Information Extraction.}
The aggregated textual content is then passed to an information-extraction agent, which identifies key semantic elements such as relevant function names, file paths, and bug commits. 
The agent is prompted to reason about the likely code locations involved in the bug based on the bug description and related contextual information. 
To retrieve the precise code regions, we equip the agent with a tool named \texttt{extract\_function}, which locates function bodies using the extracted function name, file path, and commit.

These steps collectively prepare the structured input required for downstream context retrieval and alignment synthesis. While necessary for the overall workflow, they rely on standard techniques and are not considered part of the novel contributions of \toolname{}.

\section{Evaluation on Real-world Projects}
\label{appendix:real-world}

\subsection{Setup and Metrics.}
To evaluate the effectiveness of \toolname{}, we first align it using existing issues and then apply it to real-world C/C++ projects to detect new bugs. We focus on three types of bugs, namely Out-of-Bounds (OOB), Divide-by-Zero (DBZ), and Memory Leak (MLK), which span a wide spectrum of program properties including pointer-related, numeric, and combined properties.

\paragraph{Learning-phase Example Selection.} 
For each bug type, we select two to three existing issues from existing works for the learning phase~\cite{guo2022precise, guo2024precise, guo2025repoaudit}. We use the criterion of prioritizing cases whose causal chains span multiple functions and whose issue discussions contain more than three rounds of comments. Such cases provide richer reasoning trajectories for synthesizing robust retrieval strategies and detection guidelines. In total, seven bugs are used for learning, each representing a distinct \buggypattern{}:
\begin{itemize}
\item \textbf{Out-of-Bounds (OOB)}.
(1) Oversized offset (OSO), where the offset exceeds the buffer length;
(2) Negative offset (NOF), where the offset in a buffer operation is negative;
(3) Allocation size overflow (ASO), where integer overflow during memory allocation produces a smaller-than-expected buffer.

\item \textbf{Divide-by-Zero (DBZ)}.
(1) Insufficient zero check (IZC), where a conditional check on the divisor is insufficient to exclude zero;
(2) Literal zero division (LZD), where a literal zero value is directly used as a divisor without validation.

\item \textbf{Memory Leak (MLK)}.
(1) Unexecuted cleanup (UEC), where cleanup code exists but is skipped due to early returns or exceptions;
(2) Missing cleanup (MSC), where no code is provided to free allocated memory.
\end{itemize}

Details of the selected cases, including \buggypattern{}, corresponding commit, target file, and bug report link are shown in Table~\ref{tab:leanring-cases}.

\begin{table*}[htbp]
\centering
\caption{The details of cases used for the learning phase.}
\label{tab:leanring-cases}
\resizebox{0.8\textwidth}{!}{
\begin{tabular}{l|cc|c|l|c}
\toprule
\textbf{Project} & \textbf{Bug Type}  & \textbf{\Buggypattern{}} & \textbf{Commit} & \textbf{Target File}& \textbf{Report} \\
\midrule
frr & OOB  &OSO & 26b2fbf & zebra/kernel\_netlink.c & \href{https://github.com/FRRouting/frr/issues/11624}{link} \\
zstd & OOB  & NOF & e5db7c9 & programs/util.c & \href{https://github.com/facebook/zstd/issues/3200}{link} \\
systemd & OOB  &ASO & fa2ba7a & src/libsystemd-network/disc-router.c & \href{https://github.com/systemd/systemd/issues/23258}{link} \\

\midrule
linux & DBZ  &IZC & 9e9b451 & block/blk-mq-cpumap.c & \href{https://lore.kernel.org/linux-block/21cb65d1-b91a-2627-3824-292de3a7553a@suse.de/T/\#t}{link} \\
linux & DBZ  &LZD & 9e9b451 & drivers/video/logo/pnmtologo.c & \href{https://lore.kernel.org/linux-parisc/alpine.DEB.2.22.394.2105121353530.1204552@ramsan.of.borg/T/\#t}{link} \\

\midrule
memcached & MLK  & UEC & 6fb5ef7 & restart.c & \href{https://github.com/memcached/memcached/pull/1216}{link} \\
libuv & MLK  &MSC& 98a4bab & docs/code/plugin/main.c & \href{https://github.com/libuv/libuv/pull/4720}{link} \\
\bottomrule
\end{tabular}
}
\vspace{-3mm}
\end{table*}

\paragraph{Project Selection.}
We selected six high-profile C/C++ projects on GitHub following two criteria:
(1) large-scale codebases (over 100K LoC), and 
(2) active maintenance (substantial development activity within the past six months).
These six projects are representative real-world systems, averaging 654K LoC and 22.3K GitHub stars.

\paragraph{Baselines.} To evaluate the effectiveness of \toolname{}, we compare it against the state-of-the-art LLM-driven bug detection tool RepoAudit~\citep{guo2025repoaudit}, as well as three commercial tools, namely BugBot~\citep{cursorbugbot}, \citet{coderabbit}, and Infer~\citep{Infer}. RepoAudit is an LLM agent that detects data-flow bugs in repositories. BugBot and CodeRabbit are commercial agents that can be integrated with GitHub repositories to automatically scan pull requests. Infer is a production-grade static analyzer with a fixed set of built-in checkers.

\paragraph{Evaluation Settings.}
For each project, we applied the synthesized retrieval strategies and detection prompts to detect new bugs. To control computational cost, we randomly sampled 100 seed statements per bug type per repository (1800 seeds in total). All baselines were evaluated using the same seed statements and project repositories to ensure a fair comparison. We use Claude 3.7 to power RepoAudit, the same as \toolname{}. Since ground-truth labels for recall are unavailable in this real-world setting, we report precision only. Each detected bug report was independently examined by two authors, with disagreements adjudicated by a third author. Only cases where all three reviewers reached consensus were considered true positives (TP) and subsequently reported to developers.

\subsection{Result}
As shown in Table~\ref{table:new_bug}, \toolname{} identifies 8 DBZ, 8 OOB, and 64 MLK bugs, yielding an overall precision of 76.92\%. Precision is highest for MLK at 81.01\%, while DBZ and OOB achieve 66.67\% and 61.54\% respectively, largely due to the limited backward control-flow and data-flow context available from the extracted seeds.

A comparative analysis with existing baselines is presented in Table~\ref{tab:real-world-compare}.
RepoAudit performs comparably to \toolname{} on MLK bugs, but it identifies only one true bug across OOB and DBZ while producing 37 false positives, resulting in a substantially lower overall precision of 53.64\%.
BugBot and CodeRabbit detect only 3 and 12 bugs, achieving precisions of 50.00\% and 48.00\%, respectively.
Infer produces no valid reports for OOB or DBZ and detects only 6 MLK bugs, with a precision of 22.22\%.
Across all six real-world projects, \toolname{} consistently surpasses these baselines in both the number of detected issues and overall precision, demonstrating strong robustness and generality across large and diverse codebases.

\begin{table*}[t]
    \centering
    \small
    \caption{The comparison results between \toolname{} and baselines on 6 real-world projects.}
    \label{tab:real-world-compare}
    \resizebox{0.9\textwidth}{!}{%
    \begin{tabular}{l|ccc|ccc|ccc|ccc}
    \toprule
    \multirow{2}{*}{\textbf{Tool}}
        & \multicolumn{3}{c|}{\textbf{OOB}}
        & \multicolumn{3}{c|}{\textbf{DBZ}}
        & \multicolumn{3}{c|}{\textbf{MLK}}
        & \multicolumn{3}{c}{\textbf{Total}} \\
    \cmidrule(lr){2-4} \cmidrule(lr){5-7} \cmidrule(lr){8-10} \cmidrule(lr){11-13}
        & \textbf{\# TP} & \textbf{\# FP} & \textbf{P(\%)}
        & \textbf{\# TP} & \textbf{\# FP} & \textbf{P(\%)}
        & \textbf{\# TP} & \textbf{\# FP} & \textbf{P(\%)}
        & \textbf{\# TP} & \textbf{\# FP} & \textbf{P(\%)}
        \\
    \midrule
    BugScope   & 8 & 5 & 61.54 & 8 & 4 & 66.67 & 64 & 15 & 81.01 & 80 & 24 & 76.92 \\
    RepoAudit  & 1 & 13 & 7.14  & 0 & 24 & 0.00  & 58 & 14 & 80.56 & 59 & 51 & 53.64 \\
    BugBot     & 0 & 2 & 0.00   & 0 & 0 & 0.00   & 3 & 1 & 75.00  & 3 & 3 & 50.00  \\
    CodeRabbit & 1 & 8 & 11.11  & 0 & 0 & 0.00   & 11 & 5 & 68.75 & 12 & 13 & 48.00 \\
    Infer      & 0 & 0 & 0.00   & 0 & 0 & 0.00   & 6 & 21 & 22.22 & 6 & 21 & 22.22 \\
    \bottomrule
    \end{tabular}
    }
\end{table*}

\section{Details of the Dataset}
\label{appendix:dataset}

\begin{table*}[htbp]
\centering
\vspace{-4mm}
\caption{The details of the evaluation dataset.}
\label{tab:dataset-details}
\resizebox{\textwidth}{!}{
\begin{tabular}{l|cc|c|l|c}
\toprule
\textbf{Project} & \textbf{Bug Type}  & \textbf{\Buggypattern{}} & \textbf{Commit} & \textbf{Target File}& \textbf{Report} \\
\midrule
redis & OOB  &OSO & 93dda65 & src/t\_zset.c & \href{https://github.com/redis/redis/issues/10462}{link} \\
zstd & OOB  & OSO & e5db7c9 & programs/util.c & \href{https://github.com/facebook/zstd/pull/3300}{link} \\
systemd & OOB  &OSO & fa2ba7a & src/basic/time-util.c & \href{https://github.com/systemd/systemd/issues/23928}{link} \\
qemu & OOB  &OSO & 232e925 & contrib/elf2dmp/qemu\_elf.c & \href{https://gitlab.com/qemu-project/qemu/-/issues/1013}{link} \\

curl & OOB  &NOF & e0c68f02 & lib/sendf.c & \href{https://github.com/curl/curl/issues/9149}{link} \\
curl & OOB  &NOF & e0c68f02 & lib/sendf.c & \href{https://github.com/curl/curl/issues/9149}{link} \\
php-src & OOB  &NOF & 492f9c6 & ext/opcache/zend\_accelerator\_blacklist.c & \href{https://github.com/php/php-src/issues/9033}{link} \\
openssl & OOB  &NOF & 2837b19 & crypto/bf/bf\_ofb64.c & \href{https://github.com/openssl/openssl/issues/18751}{link} \\

php-src & OOB  &ASO & 492f9c6 & sapi/cli/php\_cli\_server.c & \href{https://github.com/php/php-src/issues/8989}{link} \\
frr & OOB  &ASO & 26b2fbf & bfdd/control.c & \href{https://github.com/FRRouting/frr/issues/11097}{link} \\
binutils-gdb & OOB  &ASO & 4bb461e & ld/libdep\_plugin.c & \href{https://sourceware.org/bugzilla/show_bug.cgi?id=29101}{link} \\
gcc & OOB  &ASO & 9715f10 & libcpp/files.cc & \href{https://gcc.gnu.org/bugzilla/show_bug.cgi?id=105403}{link} \\ 

\midrule

git & DBZ  &IZC & 49ac1d3 & git/builtin/pack-objects.c & \href{https://lore.kernel.org/git/YI1f0NesWZFqh9O2@coredump.intra.peff.net/T/\#m81221758bb89fa0088a6697f5863242388490bc2}{link} \\
binutils-gdb & DBZ  &IZC & 2005aa02 & gdb/amd64-tdep.c & \href{https://sourceware.org/bugzilla/show_bug.cgi?id=27847}{link} \\
openssl & DBZ  &IZC & bc8c3627 & crypto/pkcs12/p12\_key.c & \href{https://github.com/openssl/openssl/issues/16331}{link} \\
vim & DBZ  &IZC & ccfb7c67 & vim/src/misc2.c & \href{https://github.com/vim/vim/issues/8767}{link} \\
systemd & DBZ  &IZC & f6e40037 & src/shared/creds-util.c & \href{https://github.com/systemd/systemd/issues/20469}{link} \\
ImageMagick & DBZ  &IZC & 442c87b9 & MagickCore/cache.c & \href{https://github.com/ImageMagick/ImageMagick/issues/3642}{link} \\
ImageMagick & DBZ  &IZC & 442c87b9 & MagickCore/cache.c & \href{https://github.com/ImageMagick/ImageMagick/issues/3653}{link} \\
libuv & DBZ  &IZC& af1a79cf & src/unix/linux-core.c & \href{https://github.com/libuv/libuv/pull/3166/commits/09fa971023e4139a9f4e6c3356959de01476a605}{link} \\ 

linux & DBZ  &LZD & 9e9b451 & lib/math/rational.c & \href{https://lore.kernel.org/lkml/CAM7=BFoH7Q+YHvPFnHM4j72ORHQp4gTjHFjnfeLsV2-30ZLNYw@mail.gmail.com/T/\#t}{link} \\
linux & DBZ  &LZD & 9e9b451 & drivers/char/agp/isoch.c & \href{https://lore.kernel.org/dri-devel/YJ00C\%2FKdhe3bSrtH@kroah.com/T/\#u}{link} \\
goaccess & DBZ  &LZD & 0abddd5f & src/gholder.c & \href{https://github.com/allinurl/goaccess/issues/2106}{link} \\
goaccess & DBZ  &LZD & 0abddd5f & src/gholder.c & \href{https://github.com/allinurl/goaccess/issues/2106}{link} \\

\midrule
memcached & MLK  & UEC & e15e1d6 & memcached.c & \href{https://github.com/Memcached/Memcached/issues/337}{link} \\
libsass & MLK  & UEC & 4da7c4b & src/permutate.hpp & \href{https://github.com/sass/libsass/issues/3014}{link} \\
h3 & MLK  & UEC & 3a02395 & src/apps/filters/h3.c & \href{https://github.com/uber/h3/commit/2f689a3948d86b4107050d6f407216639e2716e9}{link} \\
TrinityEmulator & MLK  &UEC & ad25460 & contrib/elf2dmp/main.c & \href{https://github.com/TrinityEmulator/TrinityEmulator/commit/a570269885e296d22e58b14a4a8b100775679b9b}{link} \\

rtl\_433 & MLK  &MSC& 474feb5 & rtl\_433/src/sdr.c & \href{https://github.com/merbanan/rtl_433/pull/3202}{link} \\
linux & MLK  & MSC & 4cd8371 & drivers/net/nfpcore/nfp\_cppcore.c & \href{https://lore.kernel.org/all/20211209061511.122535-1-niejianglei2021@163.com/}{link} \\
linux & MLK  &MSC& 73b73bac & mm/damon/reclaim.c & \href{https://lore.kernel.org/all/20220714063746.2343549-1-niejianglei2021@163.com/}{link} \\
TrinityEmulator & MLK  &MSC & ad25460 & contrib/elf2dmp/main.c & \href{https://github.com/TrinityEmulator/TrinityEmulator/commit/501593c3a71a0661c4913d719ddccee2f60d0c90}{link} \\
binutils-gdb & MLK  &MSC & 0ebc886 & binutils/bucomm.c & \href{https://sourceware.org/git/gitweb.cgi?p=binutils-gdb.git;h=d6e1d48c83b165c129cb0aa78905f7ca80a1f682}{link} \\
\bottomrule
\end{tabular}
}
\vspace{-3mm}
\end{table*}

To ensure correctness and high-quality ground truth, we collected all bugs from prior works that include manually validated root causes, including 12 Divide-by-Zero cases from~\citep{guo2022precise}, 12 Out-of-Bounds cases from~\citep{guo2024precise}, and 9 Memory Leak cases from~\citep{guo2025repoaudit}. For each case, we manually examined the corresponding issue discussions and code history and labeled the exact root-cause file and context.  Compared with existing large-scale datasets that annotate bug root causes primarily through diff-patch commits~\cite{risse2025top}, our careful validation ensures that every case has a reliable and unambiguous root-cause annotation, which is essential for evaluating the quality of end-to-end bug reports. Since each bug type encompasses multiple \buggypatterns{}, we further categorize the cases according to the guidelines synthesized in the learning stage. This categorization allows us to evaluate the performance of different detection tools on specific \buggypatterns{}. Table~\ref{tab:dataset-details} presents the detailed information of our dataset.

For projects with particularly large codebases, we select a relevant subdirectory that encompasses the complete bug trace as the analysis target. The dataset comprises 33 cases drawn from 20 open-source projects, averaging 81K lines of code and 29K GitHub stars. Each case is annotated with the target file containing the bug and linked to the original bug report, facilitating precise localization and reliable evaluation.

\section{Evaluation with More Reasoning Models}
\label{appendix:moreLLMs}
To assess the robustness of \toolname{} across different reasoning engines, we further evaluate it with DeepSeek-R1 and OpenAI o4-mini in addition to Claude 3.7. The results are summarized in Table~\ref{tab:differentLLM}. When powered by Claude 3.7, \toolname{} achieves the best performance, reproducing 29 of 33 bugs, discovering 8 additional ones, and reaching 86.05\% precision, 87.88\% recall, and an F1 score of 0.87. Both DeepSeek-R1 and o4-mini maintain strong performance, achieving F1 scores of 0.84.
In terms of efficiency, the average detection cost per seed is significantly lower with DeepSeek-R1 and o4-mini (about \$0.03) than with Claude 3.7 (\$0.12). These results suggest that while Claude 3.7 provides the best overall accuracy, DeepSeek-R1 and o4-mini offer more cost-effective alternatives with competitive precision and recall.

\begin{table*}[htbp]
\centering
\vspace{-4mm}
\caption{Statistics of \toolname{} with 
\includegraphics[height=1em]{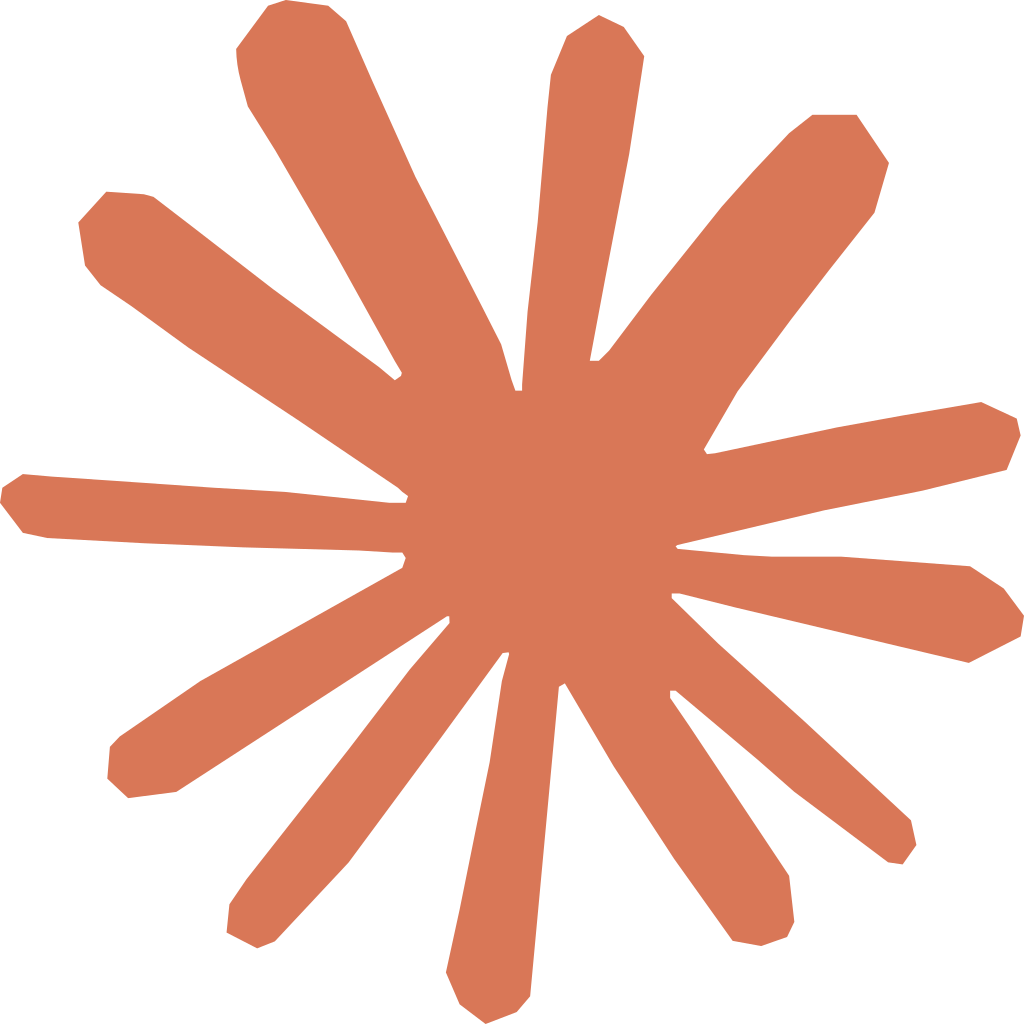}Claude 3.7 Sonnet Thinking,
\includegraphics[height=1em]{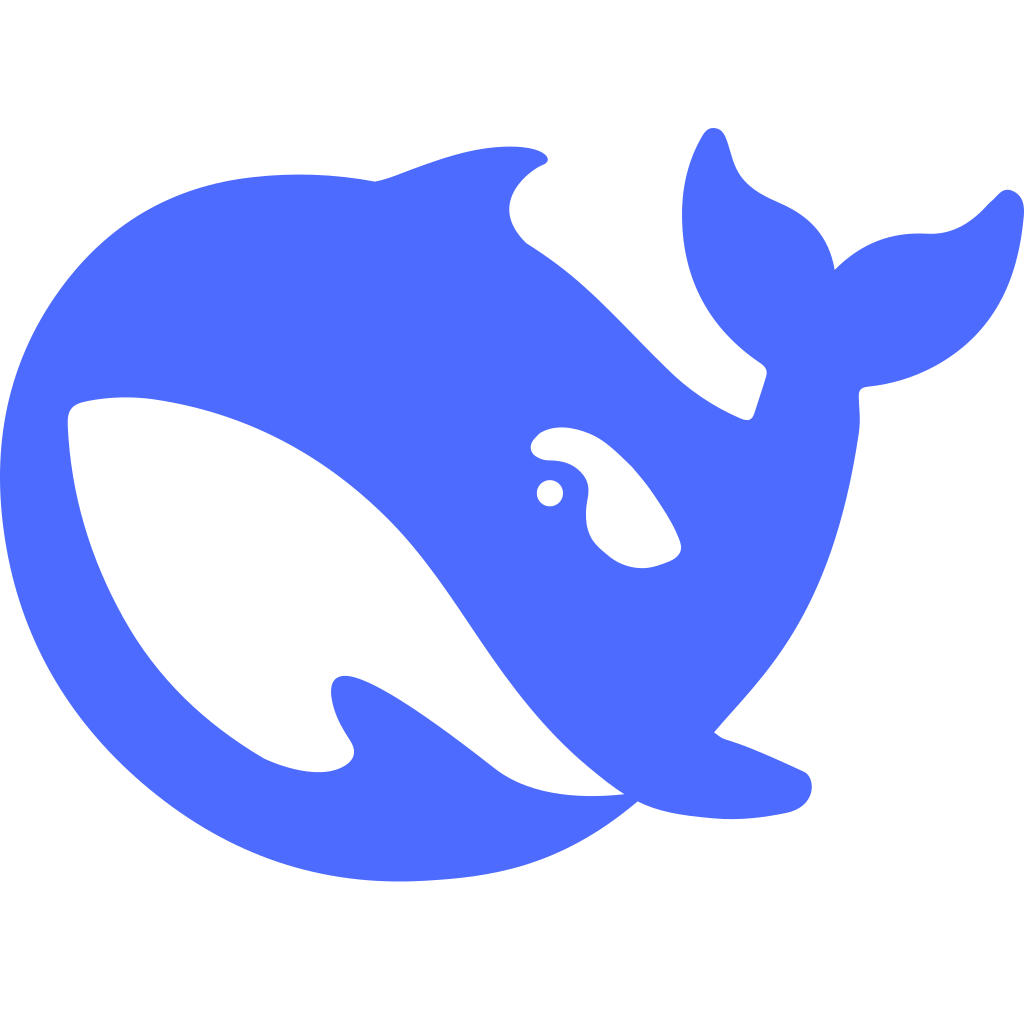}DeepSeek-R1, and 
\includegraphics[height=1em]{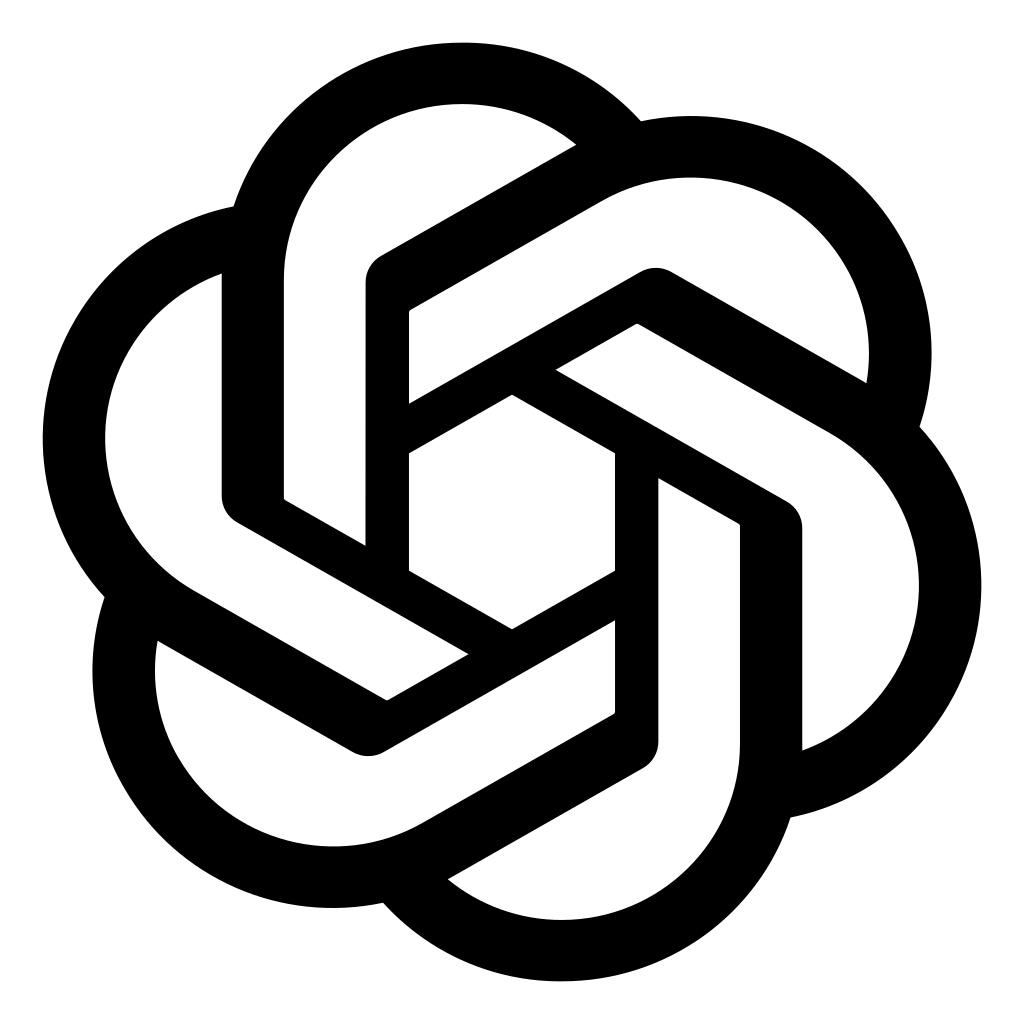}OpenAI o4-mini. 
\textbf{AP}: Anti-Pattern. 
\textbf{\#C}: number of cases. 
\textbf{\#Seed}: number of extracted seeds. 
\textbf{\#R}: number of reproduced bugs. 
\textbf{\#N}: number of new bugs. 
\textbf{P(\%)}, \textbf{R(\%)}, \textbf{F1}: precision, recall, F1 score. 
\textbf{Time}, \textbf{Cost}: total analysis time and financial cost. \xz{I cannot tell which APs belong to which bug type} \gjy{Resolved.}}
\label{tab:differentLLM}
\resizebox{\textwidth}{!}{
\begin{tabular}{l|c|c|ccc|cccc|ccc|cc}
\toprule
\textbf{Model} & \textbf{Type} & \textbf{AP} & \textbf{\#C} & \textbf{\#Seed} & \textbf{\#R} & \textbf{\#N} & \textbf{\#TP} & \textbf{\#FP} & \textbf{P(\%)} & \textbf{R(\%)} & \textbf{F1} & \textbf{Time (s)} & \textbf{Cost (\$)} \\
\midrule
\multirow{8}{*}{\includegraphics[height=2em]{Fig/claude.png}}
 & \multirow{3}{*}{OOB} & OSO & 4 & 269 & 3 & 1 & 4 & 2 & 66.67 & 75.00 & 0.71 & 5,547 & 32.73 \\
 &  & NOF & 4 & 161 & 4 & 0 & 4 & 2 & 66.67 & 100.00 & 0.80 & 2,567 & 12.35 \\
 &  & ASO & 4 & 27  & 4 & 0 & 4 & 0 & 100.00 & 100.00 & 1.00 & 1,700 & 4.78 \\
 \cmidrule{2-3}
 & \multirow{2}{*}{DBZ} & IZC & 8 & 32  & 5 & 1 & 6 & 1 & 85.71 & 62.50 & 0.72 & 1,548 & 4.24 \\
 &  & LZD & 4 & 16  & 4 & 5 & 9 & 0 & 100.00 & 100.00 & 1.00 & 653   & 1.19 \\
 \cmidrule{2-3}
 & \multirow{2}{*}{MLK} & UEC & 4 & 31  & 4 & 1 & 5 & 1 & 83.33 & 100.00 & 0.91 & 1,523 & 5.83 \\
 &  & MSC & 5 & 10  & 5 & 0 & 5 & 0 & 100.00 & 100.00 & 1.00 & 1,910 & 2.24 \\
\cmidrule{2-14}
 & \multicolumn{2}{c|}{Total} & 33 & 546 & 29 & 8 & 37 & 6 & \textbf{86.05} & \textbf{87.88} & \textbf{0.87} & 15,448 & 63.36 \\
\midrule
\multirow{8}{*}{\includegraphics[height=2em]{Fig/deepseek.png}}
 & \multirow{3}{*}{OOB} & OSO & 4 & 269 & 2 & 1 & 3 & 1 & 75.00 & 50.00 & 0.60 & 12,036 & 6.06 \\
 &  & NOF & 4 & 161 & 4 & 0 & 4 & 1 & 80.00 & 100.00 & 0.89 & 2,552 & 1.56 \\
 &  & ASO & 4 & 27  & 4 & 0 & 4 & 0 & 100.00 & 100.00 & 1.00 & 1,362 & 0.80 \\
 \cmidrule{2-3}
 & \multirow{2}{*}{DBZ} & IZC & 8 & 32  & 4 & 1 & 5 & 0 & 100.00 & 50.00 & 0.67 & 1,548 & 4.24 \\
 &  & LZD & 4 & 16  & 4 & 3 & 7 & 1 & 87.50 & 100.00 & 0.93 & 653   & 1.19 \\
 \cmidrule{2-3}
 & \multirow{2}{*}{MLK} & UEC & 4 & 31  & 4 & 1 & 5 & 1 & 83.33 & 100.00 & 0.91 & 1,166 & 0.79 \\
 &  & MSC & 5 & 10  & 4 & 0 & 4 & 0 & 100.00 & 80.00 & 0.89 & 2,358 & 0.83 \\
\cmidrule{2-14}
 & \multicolumn{2}{c|}{Total} & 33 & 546 & 26 & 6 & 32 & 4 & \textbf{88.89} & \textbf{78.79} & \textbf{0.84} & 21,675 & 15.48 \\
\midrule
\multirow{8}{*}{\includegraphics[height=2em]{Fig/openai.png}}
 & \multirow{3}{*}{OOB} & OSO & 4 & 269 & 3 & 0 & 3 & 2 & 60.00 & 75.00 & 0.67 & 9,161 & 7.83 \\
 &  & NOF & 4 & 161 & 4 & 0 & 4 & 1 & 80.00 & 100.00 & 0.89 & 1,615 & 2.63 \\
 &  & ASO & 4 & 27  & 4 & 0 & 4 & 0 & 100.00 & 100.00 & 1.00 & 803   & 1.01 \\
 \cmidrule{2-3}
 & \multirow{2}{*}{DBZ} & IZC & 8 & 32  & 4 & 0 & 4 & 0 & 100.00 & 50.00 & 0.67 & 1,146 & 1.83 \\
 &  & LZD & 4 & 16  & 4 & 2 & 6 & 0 & 100.00 & 100.00 & 1.00 & 371   & 0.49 \\
 \cmidrule{2-3}
 & \multirow{2}{*}{MLK} & UEC & 4 & 31  & 4 & 1 & 5 & 2 & 71.43 & 100.00 & 0.83 & 648   & 0.97 \\
 &  & MSC & 5 & 10  & 4 & 1 & 5 & 0 & 100.00 & 80.00 & 0.89 & 752   & 0.67 \\
\cmidrule{2-14}
 & \multicolumn{2}{c|}{Total} & 33 & 546 & 27 & 4 & 31 & 5 & \textbf{86.11} & \textbf{81.82} & \textbf{0.84} & 14,496 & 15.43 \\
\bottomrule
\end{tabular}
}
\end{table*}

\section{Detailed Comparison Result with Baselines}
\label{appendix:baseline}

\paragraph{Setup and Metrics.}
We evaluate all baselines on the dataset described in Appendix~\ref{appendix:dataset}. For fairness, we restrict the comparison to cases included in the evaluation set, and use Claude~3.7, a strong reasoning model (see Section~\ref{sec:reproduce}), to power both Claude Code and RepoAudit.
For Claude Code, we provide the seven examples used in the learning stage and prompt it to analyze the code and synthesize detection guidelines, which are then applied to detect bugs in the target file. We also equip Claude Code with tools such as Grep, Glob, and Read to support context retrieval. 
To further examine the impact of stronger foundation models on agent-based bug detection, we additionally evaluate Claude Code powered by Claude Opus~4.6, a state-of-the-art model for code analysis. The comparison between the two configurations is shown in Table~\ref{tab:claude_compare}.
For RepoAudit, we extend the system with source and sink extractors for DBZ and OOB, and similarly restrict the analysis to the file containing the original bug.
For BugBot and CodeRabbit, we simulate GitHub pull requests containing the target file and collect reported issues that match the corresponding bug type. Since Infer is a compiler-based tool, we run it from the project root and only consider reports relevant to the target bug type within the corresponding file.

We report precision, recall, F1 score, and the number of newly discovered bugs.

\paragraph{Results.}
As shown in Table~\ref{tab:baseline} and Table~\ref{tab:claude_compare}, \toolname{} significantly outperforms all baselines across OOB, DBZ, and MLK categories. Claude Code achieves 65.38\% precision and 42.42\% recall, much lower than \toolname{}. Although BugBot and CodeRabbit attain comparable precision, their recall remains poor, leading to low F1 scores of 0.43 and 0.29. RepoAudit demonstrates strong recall on MLK bugs but suffers from low precision on DBZ and OOB, yielding an overall F1 of 0.38. Infer performs the worst, detecting only one OOB bug with an F1 of 0.05.

Claude Code powered by Claude Opus~4.6 achieves the highest F1 score of 0.51 among all baselines. By leveraging tools such as Grep, it is able to effectively track dependencies and retrieve relevant context. However, the detection guidelines it synthesizes tend to overfit to code patterns in the provided examples and lack a structured reasoning framework. As a result, its analysis is often verbose but unfocused, frequently failing to recognize existing guards as sufficient.
In contrast, \toolname{} decomposes bug detection into three stages and enforces structured reasoning chains inspired by human auditing practices, achieving substantially higher precision and recall.

RepoAudit adopts a source–sink paradigm tailored to data-flow vulnerabilities. While this supports customization of sources and sinks, many DBZ and OOB bugs cannot be expressed as data-flow reachability problems, which require reasoning about numerical relationships and multi-domain properties. Consequently, RepoAudit achieves high F1 on unexecuted cleanup (UEC) and missing cleanup (MSC) but fails on patterns such as negative offset (NOF) and insufficient zero check (IZC).


BugBot and CodeRabbit also struggle with complex \buggypatterns{}. They rely on LLM prior knowledge to detect common issues but fail on intricate patterns such as NOF, ASO, and IZC. Moreover, both suffer from insufficient context retrieval. BugBot supports limited inter-procedural reasoning within a single file but cannot trace bug propagation across files, leading to poor coverage and many false negatives. CodeRabbit exhibits even weaker retrieval, and its explanations are overly generic. For example, among its two true positives of literal zero division (LZD), the explanation only states that ``if the divisor is zero, a divide-by-zero error may occur,'' without justifying why the divisor can be zero. In contrast, \toolname{} reconstructs the full propagation path and provides auditor-style explanations. Case studies of BugBot and CodeRabbit are presented in Appendix~\ref{appendix:case_study}.

Infer further suffers from build failures and limited checkers. We were able to evaluate only 22 of 40 cases; others failed due to incompatibilities with Infer’s build interception (\texttt{infer capture}) under GCC-specific flags, custom Makefiles, or non-standard toolchains. Even among successful builds, Infer detected only one true positive and six false positives. Notably, it has no checker for DBZ, resulting in zero reports for this category. For OOB, although related checkers exist, Infer cannot distinguish between OSO and NOF, limiting its effectiveness.

In contrast, \toolname{} replicates the human process of learning \buggypatterns{}. By leveraging the generalization ability of LLM-based reasoning and the synthesized detection prompt, it adapts to diverse bug types and consistently delivers high precision, recall, and informative explanations, even where traditional or commercial tools fail.

\begin{table*}[htbp]
\centering
\caption{The comparison results between \toolname{} and baselines. 
\textbf{AP} indicates \textbf{Anti-Pattern}.
\textbf{\#C}: number of cases. \textbf{\#R}: reproduced bugs. \textbf{\#N}: new bugs. 
\textbf{\#TP} / \textbf{\#FP}: true / false positives. 
\textbf{P(\%)}, \textbf{R(\%)}, \textbf{F1}: precision, recall, F1 score.}
\label{tab:baseline}
\resizebox{0.85\textwidth}{!}{
\begin{tabular}{c|c|c|c|cccc|ccc}
\toprule
\textbf{Tool Name} & \textbf{Type} & \textbf{AP} & \textbf{\#C} 
& \textbf{\#R} & \textbf{\#N} & \textbf{\#TP} & \textbf{\#FP} 
& \textbf{P(\%)} & \textbf{R(\%)} & \textbf{F1} \\
\midrule
\multirow{7}{*}{\textbf{BugScope}} 
 & \multirow{3}{*}{OOB} & OSO & 4 & 3 & 1 & 4 & 2 & 66.67 & 75.00 & 0.71 \\
 &  & NOF & 4 & 4 & 0 & 4 & 2 & 66.67 & 100.00 & 0.80 \\
 &  & ASO & 4 & 4 & 0 & 4 & 0 & 100.00 & 100.00 & 1.00 \\
 \cmidrule{2-3}
 & \multirow{2}{*}{DBZ} & IZC & 8 & 5 & 1 & 6 & 1 & 85.71 & 62.50 & 0.72 \\
 &  & LZD & 4 & 4 & 5 & 9 & 0 & 100.00 & 100.00 & 1.00 \\
 \cmidrule{2-3}
 & \multirow{2}{*}{MLK} & UEC & 4 & 4 & 1 & 5 & 1 & 83.33 & 100.00 & 0.91 \\
 &  & MSC & 5 & 5 & 0 & 5 & 0 & 100.00 & 100.00 & 1.00 \\ \cmidrule{2-11}
 & \multicolumn{2}{c|}{Total} & 33 & 29 & 8 & 37 & 6 & \textbf{86.05} & \textbf{87.88} & \textbf{0.87} \\

\midrule

\multirow{7}{*}{\textbf{RepoAudit}} 
 & \multirow{3}{*}{OOB} & OSO & 4 & 1 & 0 & 1 & 4 & 20.00 & 25.00 & 0.22 \\
 &  & NOF & 4 & 0 & 0 & 0 & 1 & 0.00 & 0.00 & 0.00 \\
 &  & ASO & 4 & 1 & 0 & 1 & 2 & 33.33 & 25.00 & 0.29 \\
 \cmidrule{2-3}
 & \multirow{2}{*}{DBZ} & IZC & 8 & 1 & 0 & 1 & 14 & 6.67 & 12.50 & 0.09 \\
 &  & LZD & 4 & 3 & 0 & 3 & 6 & 33.33 & 75.00 & 0.46 \\
 \cmidrule{2-3}
 & \multirow{2}{*}{MLK} & UEC & 4 & 4 & 1 & 5 & 1 & 83.33 & 100.00 & 0.91 \\
 &  & MSC & 5 & 4 & 0 & 4 & 0 & 100.00 & 80.00 & 0.89 \\ \cmidrule{2-11}
 & \multicolumn{2}{c|}{Total} & 33 & 14 & 1 & 15 & 28 & \textbf{34.88} & \textbf{42.42} & \textbf{0.38} \\
\midrule

\multirow{7}{*}{\textbf{BugBot}} 
 & \multirow{3}{*}{OOB} & OSO & 4 & 1 & 1 & 2 & 1 & 66.67 & 25.00 & 0.36 \\
 &  & NOF & 4 & 1 & 0 & 1 & 2 & 33.33 & 25.00 & 0.29 \\
 &  & ASO & 4 & 0 & 1 & 1 & 1 & 50.00 & 0.00 & 0.00 \\
 \cmidrule{2-3}
 & \multirow{2}{*}{DBZ} & IZC & 8 & 1 & 0 & 1 & 0 & 100.00 & 12.50 & 0.22 \\
 &  & LZD & 4 & 3 & 1 & 4 & 0 & 100.00 & 75.00 & 0.86 \\
 \cmidrule{2-3}
 & \multirow{2}{*}{MLK} & UEC & 4 & 1 & 1 & 2 & 2 & 50.00 & 25.00 & 0.33 \\
 &  & MSC & 5 & 3 & 1 & 4 & 0 & 100.00 & 60.00 & 0.75 \\ \cmidrule{2-11}
 & \multicolumn{2}{c|}{Total} & 33 & 10 & 5 & 15 & 6 & \textbf{71.43} & \textbf{30.30} & \textbf{0.43} \\
\midrule

\multirow{7}{*}{\textbf{CodeRabbit}} 
 & \multirow{3}{*}{OOB} & OSO & 4 & 0 & 0 & 0 & 3 & 0.00 & 0.00 & 0.00 \\
 &  & NOF & 4 & 0 & 0 & 0 & 0 & 0.00 & 0.00 & 0.00 \\
 &  & ASO & 4 & 1 & 0 & 1 & 0 & 100.00 & 25.00 & 0.40 \\
 \cmidrule{2-3}
 & \multirow{2}{*}{DBZ} & IZC & 8 & 0 & 0 & 0 & 0 & 0.00 & 0.00 & 0.00 \\
 &  & LZD & 4 & 1 & 1 & 2 & 0 & 100.00 & 25.00 & 0.40 \\
 \cmidrule{2-3}
 & \multirow{2}{*}{MLK} & UEC & 4 & 0 & 0 & 0 & 0 & 0.00 & 0.00 & 0.00 \\
 &  & MSC & 5 & 4 & 0 & 4 & 0 & 100.00 & 80.00 & 0.89 \\ \cmidrule{2-11}
 & \multicolumn{2}{c|}{Total} & 33 & 6 & 1 & 7 & 3 & \textbf{70.00} & \textbf{18.18} & \textbf{0.29} \\
\midrule

\multirow{7}{*}{\textbf{Infer}} 
 & \multirow{3}{*}{OOB} & OSO & 4 & 0 & 0 & 0 & 6 & 0.00 & 0.00 & 0.00 \\
 &  & NOF & 4 & 1 & 0 & 1 & 1 & 50.00 & 25.00 & 0.33 \\
 &  & ASO & 4 & 0 & 0 & 0 & 0 & 0.00 & 0.00 & 0.00 \\
 \cmidrule{2-3}
 & \multirow{2}{*}{DBZ} & IZC & 8 & 0 & 0 & 0 & 0 & 0.00 & 0.00 & 0.00 \\
 &  & LZD & 4 & 0 & 0 & 0 & 0 & 0.00 & 0.00 & 0.00 \\
 \cmidrule{2-3}
 & \multirow{2}{*}{MLK} & UEC & 4 & 0 & 0 & 0 & 0 & 0.00 & 0.00 & 0.00 \\
 &  & MSC & 5 & 0 & 0 & 0 & 0 & 0.00 & 0.00 & 0.00 \\ \cmidrule{2-11}
 & \multicolumn{2}{c|}{Total} & 33 & 1 & 0 & 1 & 7 & \textbf{12.50} & \textbf{3.03} & \textbf{0.05} \\
\bottomrule
\end{tabular}
}
\end{table*}

\begin{table}[t]
\centering
\caption{The result of Claude Code powered by Claude Sonnet 3.7 and Claude Opus 4.6.
\textbf{AP} indicates \textbf{Anti-Pattern}.
\textbf{\#C}: number of cases. \textbf{\#R}: reproduced bugs. \textbf{\#N}: new bugs. 
\textbf{\#TP} / \textbf{\#FP}: true / false positives. 
\textbf{P(\%)}, \textbf{R(\%)}, \textbf{F1}: precision, recall, F1 score.
}
\label{tab:claude_compare}
\resizebox{0.75\linewidth}{!}{
\begin{tabular}{c|c|c|c|cccc|ccc}
\toprule
\textbf{Model} & \textbf{Type} & \textbf{AP} & \textbf{\#C} 
& \textbf{\#R} & \textbf{\#N} & \textbf{\#TP} & \textbf{\#FP} 
& \textbf{P(\%)} & \textbf{R(\%)} & \textbf{F1} \\
\midrule

\multirow{7}{*}{\textbf{Sonnet 3.7}} 
 & \multirow{3}{*}{OOB} & OSO & 4 & 1 & 0 & 1 & 1 & 50.00 & 25.00 & 0.33 \\
 &  & NOF & 4 & 0 & 0 & 0 & 3 & 0.00 & 0.00 & 0.00 \\
 &  & ASO & 4 & 2 & 0 & 2 & 1 & 66.67 & 50.00 & 0.57 \\
 \cmidrule{2-3}
 & \multirow{2}{*}{DBZ} & IZC & 8 & 3 & 0 & 3 & 5 & 37.50 & 37.50 & 0.38 \\
 &  & LZD & 4 & 3 & 1 & 4 & 1 & 80.00 & 75.00 & 0.77 \\
 \cmidrule{2-3}
 & \multirow{2}{*}{MLK} & UEC & 4 & 1 & 1 & 2 & 0 & 100.00 & 25.00 & 0.40 \\
 &  & MSC & 5 & 3 & 1 & 4 & 1 & 80.00 & 60.00 & 0.69 \\ \cmidrule{2-11}
 & \multicolumn{2}{c|}{Total} & 33 & 13 & 3 & 16 & 12 & \textbf{57.14} & \textbf{39.39} & \textbf{0.47} \\

\midrule

\multirow{7}{*}{\textbf{Opus 4.6}} 
 & \multirow{3}{*}{OOB} & BOF & 4 & 1 & 1 & 1 & 1 & 50.00 & 25.00 & 0.33 \\
 &  & BUF & 4 & 0 & 0 & 0 & 0 & 0.00 & 0.00 & 0.00 \\
 &  & ASO & 4 & 2 & 2 & 2 & 2 & 50.00 & 50.00 & 0.50 \\
 \cmidrule{2-3}
 & \multirow{2}{*}{DBZ} & IZC & 8 & 4 & 0 & 4 & 2 & 66.67 & 50.00 & 0.57 \\
 &  & LZD & 4 & 4 & 0 & 4 & 2 & 66.67 & 100.00 & 0.80 \\
 \cmidrule{2-3}
 & \multirow{2}{*}{MLK} & CNE & 4 & 1 & 1 & 2 & 0 & 100.00 & 25.00 & 0.40 \\
 &  & NCC & 5 & 2 & 2 & 4 & 2 & 66.67 & 40.00 & 0.50 \\ \cmidrule{2-11}
 & \multicolumn{2}{c|}{Total} & 33 & 14 & 6 & 17 & 9 & \textbf{65.38} & \textbf{42.42} & \textbf{0.51} \\

\bottomrule
\end{tabular}
}
\end{table}

\paragraph{Financial and Computational Cost.}
We also compare the monetary and computational cost of \toolname{} against baselines. 
Among the baselines, RepoAudit is the only tool that publicly reports per-query usage cost. Under the same evaluation settings, RepoAudit incurs an average cost of \$0.26 per seed, which is higher than the cost of \toolname{}. BugBot and CodeRabbit are subscription-based commercial products whose APIs do not expose 
per-project or per-seed pricing information, making a fine-grained cost comparison infeasible. 
Infer, in contrast, is a purely symbolic bug detector and therefore incurs negligible monetary cost 
aside from CPU and memory consumption, but its precision and recall are significantly lower on our benchmark, as shown in Table~\ref{tab:baseline}. 
Overall, \toolname{} achieves a favorable balance between effectiveness and cost, 
delivering substantially higher detection performance while maintaining competitive per-query expenditure.

\section{Ablation Studies}
\label{appendix:ablation}

\textbf{Setup and Metrics.} 
We evaluate two ablations, namely \emph{NoRetrieval} and \emph{NoAlign}, to assess the contributions of context retrieval and learning stage alignment, respectively. Specifically, \emph{NoRetrieval} removes the context retrieval agent and provides the bug detection agent only with the file containing the original bug. \emph{NoAlign} synthesizes prompts directly from bug reports without the data augmentation and abstraction. We report precision, recall, and F1 score for both ablations, and also track the number of newly discovered bugs.

\textbf{Results.} 
Table~\ref{tab:ablation} presents the ablation results using Claude 3.7. Without context retrieval, \emph{NoRetrieval} only discovers 3 new bugs and decreases the number of reproduced bugs by 15, reducing the recall to 42.42\%. This decline is attributed to the lack of contextual information. In real-world programs, the values that trigger bugs may propagate across multiple functions. For example, in Figure~\ref{fig:examplecode}(a), both the buffer and the index are passed from the caller function located in another file. Without visibility into the complete data-flow chain, neither human auditors nor the LLM can reliably determine the bug. To address this, we design a context retrieval agent that gathers the relevant program context and, through synthesized retrieval strategies, aligns with the reasoning process of human auditors. This alignment expands the agent’s scope and enables more precise bug detection.

When the prompt is synthesized directly from existing bug reports (\emph{NoAlign}), the number of true positives decreases to 24 while false positives increase to 38, more than six times that of \toolname{}. 
This surge in false positives arises from two main factors. First, the functions referenced in bug reports often contain large portions of code unrelated to the target \buggypattern{}, which distracts the model. Second, without data augmentation, the model-generated retrieval strategy and detection logic tend to overfit the specific code fragments described in the bug report and fail to generalize to other scenarios. For example, in the case of insufficient zero check (IZC), the true bug condition is that the divisor is bounded by a check but zero remains a possible value. Without alignment, however, the synthesized rule collapses into a shallow heuristic such as \emph{check whether the divisor is validated before division}. As a result, the detection agent simply flags any division without an explicit local check, leading to numerous false positives.

\begin{table*}[htbp]
\centering
\caption{Results of NoRetrieval and NoAlign.
\textbf{AP} indicates \textbf{Anti-Pattern}.
\textbf{\#C} denotes the number of cases for each \buggypattern{}. \textbf{\#R} denotes the number of reproduced bugs. \textbf{\#N} denotes the number of new bugs found. \textbf{\#TP} and \textbf{\#FP} are the number of true and false positives, respectively. \textbf{P(\%)} and \textbf{R(\%)} denote precision and recall. \textbf{F1} denotes the F1 score.}
\label{tab:ablation}
\resizebox{\textwidth}{!}{
\begin{tabular}{c|c|c|ccccccc|ccccccc}
\toprule
\multirow{2}{*}{\textbf{Type}} & \multirow{2}{*}{\textbf{AP}} & \multirow{2}{*}{\textbf{\#C}} 
& \multicolumn{7}{c|}{\textbf{NoRetrieval}} 
& \multicolumn{7}{c}{\textbf{NoAlign}} \\
\cmidrule{4-17}
& & & \textbf{\#R} & \textbf{\#N} & \textbf{\#TP} & \textbf{\#FP} & \textbf{P(\%)} & \textbf{R(\%)} & \textbf{F1}
  & \textbf{\#R} & \textbf{\#N} & \textbf{\#TP} & \textbf{\#FP} & \textbf{P(\%)} & \textbf{R(\%)} & \textbf{F1} \\
\midrule
\multirow{3}{*}{OOB} 
 & BOF & 4 & 0 & 1 & 1 & 0 & 100.00 & 0.00  & 0.00 & 2 & 1 & 3 & 2 & 60.00 & 50.00 & 0.55 \\
 & BUF & 4 & 1 & 0 & 1 & 2 & 33.33  & 25.00 & 0.29 & 4 & 0 & 4 & 7 & 36.36 & 100.00 & 0.53 \\
 & ASO & 4 & 3 & 0 & 3 & 1 & 75.00  & 75.00 & 0.75 & 4 & 0 & 4 & 6 & 40.00 & 100.00 & 0.57 \\
\cmidrule{1-2}
\multirow{2}{*}{DBZ} 
 & IZC & 8 & 1 & 0 & 1 & 0 & 100.00 & 12.50 & 0.22 & 2 & 0 & 2 & 5 & 28.57 & 25.00  & 0.27 \\
 & LZD & 4 & 3 & 2 & 5 & 1 & 83.33  & 75.00 & 0.79 & 2 & 2 & 4 & 1 & 80.00 & 50.00  & 0.62 \\
\cmidrule{1-2}
\multirow{2}{*}{MLK} 
 & CNE & 4 & 2 & 0 & 2 & 3 & 40.00  & 50.00 & 0.44 & 2 & 0 & 2 & 5 & 28.57 & 50.00  & 0.36 \\
 & NCC & 5 & 4 & 0 & 4 & 1 & 80.00  & 80.00 & 0.80 & 5 & 0 & 5 & 2 & 71.43 & 100.00 & 0.83 \\
\cmidrule{1-17}
\multicolumn{2}{c|}{Total} & 33 & 14 & 3 & 17 & 8 & 68.00 & 42.42 & 0.52 & 21 & 3 & 24 & 28 & 46.15 & 63.64 & 0.54 \\
\bottomrule
\end{tabular}
}
\vspace{-4mm}
\end{table*}

\section{Evaluation on Linux for Additional Bug Types}
\label{appendix:linux}

\paragraph{Setup and Metrics.}
To evaluate BugScope’s performance across a broader range of bug types, we conducted an additional controlled experiment on the Linux kernel. We used all patches from prior work~\citep{chen2025seal} as examples for the learning phase, each representing a system-specific anti-pattern, and used the synthesized retrieval strategies and detection logic to identify new bugs in the latest Linux version.

\paragraph{Results.}
As shown in Table~\ref{tab:linux-results}, we successfully uncovers 102 new bugs in the current Linux kernel with the precision of 91.07\%. These bugs span multiple categories, including Use-After-Free (UAF), Wrong Error Code (WEC), UnInitialized Variables (UIV), Null Pointer Dereference (NPD), Divide-By-Zero (DBZ), Memory-LeaK (MLK), and Out-Of-Bounds (OOB), demonstrating BugScope’s robustness across diverse bug types.

\begin{table*}[htbp]
\centering
\caption{Results of \toolname{} on Linux. \textbf{\#TP} and \textbf{\#FP} are the number of true and false positives, respectively.}
\label{tab:linux-results}
\resizebox{0.7\textwidth}{!}{
\begin{tabular}{l|c|c|c|c}
\toprule
\textbf{Bug Type} & \textbf{Original Patch} & \textbf{\# TP} & \textbf{\# FP} & \textbf{Precision (\%)} \\
\midrule
Null Pointer Dereference &
\href{https://git.kernel.org/pub/scm/linux/kernel/git/torvalds/linux.git/commit/?id=0ed554fd769a19ea8464bb83e9ac201002ef74ad}{link} &
39 & 3 & 92.86 \\
Divide by Zero &
\href{https://git.kernel.org/pub/scm/linux/kernel/git/torvalds/linux.git/commit/?id=16844e5870424c2728486dc0c0300ebf7fa09ad6}{link} &
15 & 1 & 93.75 \\
Out of Bounds &
\href{https://git.kernel.org/pub/scm/linux/kernel/git/torvalds/linux.git/commit/?id=39244cc754829bf707dccd12e2ce37510f5b1f8d}{link} &
7 & 2 & 77.78 \\
Wrong Error Code &
\href{https://git.kernel.org/pub/scm/linux/kernel/git/torvalds/linux.git/commit/?id=47c3e06ed95aa9b74932dbc6b23b544f644faf84}{link} &
20 & 2 & 90.91 \\
Use After Free &
\href{https://git.kernel.org/pub/scm/linux/kernel/git/torvalds/linux.git/commit/?id=fc7f750dc9d102c1ed7bbe4591f991e770c99033}{link} &
1 & 0 & 100.00 \\
Uninitialized Value &
\href{https://git.kernel.org/pub/scm/linux/kernel/git/torvalds/linux.git/commit/?id=41f00e6e9e55546390031996b773e7f3c1d95928}{link} &
12 & 1 & 92.31 \\
Memory Leak &
\href{https://git.kernel.org/pub/scm/linux/kernel/git/torvalds/linux.git/commit/?id=7c6daeaf0a72b4d25427df0348ad58b878a55ce3}{link} &
8 & 1 & 88.89 \\
\midrule
\textbf{Total} & --- & \textbf{102} & \textbf{10} & \textbf{91.07\%} \\
\bottomrule
\end{tabular}
}
\vspace{-3mm}
\end{table*}

\section{Case Study}
\label{appendix:case_study}

\begin{figure*}[t]
\begin{lstlisting}[
        caption={An OOB false negative example missed by Cursor BugBot and CodeRabbit.},
        label={fig:fn1}
    ]
1  static int parse_encap_seg6(struct rtattr *tb, struct in6_addr *segs){
2    struct rtattr *tb_encap[256] = {};
3	 netlink_parse_rtattr_nested(tb_encap, 256, tb);
4    ...
5    return 0;
6  }

1  void netlink_parse_rtattr_nested(struct rtattr **tb, int max, struct rtattr *rta){
2 	 netlink_parse_rtattr(tb, max, RTA_DATA(rta), RTA_PAYLOAD(rta));
3  }

1  void netlink_parse_rtattr(struct rtattr **tb, int max, struct rtattr *rta, int len){
2    memset(tb, 0, sizeof(struct rtattr *) * (max + 1));
3    ...
4  }

\end{lstlisting}
\vspace{-5mm}
\end{figure*}

\begin{figure*}[t]
\vspace{-2mm}
\begin{lstlisting}[
        caption={A DBZ false negative example missed by Cursor BugBot and CodeRabbit.},
        label={fig:fn2}
    ]
 1  static unsigned int get_number(FILE *fp) {
 2    int c, val;
 3    c = fgetc(fp);
 4    ...
 5    val = 0;
 6    while (isdigit(c)) {
 7      val = 10*val+c-'0';
 8      if (is_plain_pbm)
 9        break;
10      c = fgetc(fp);
11      if (c == EOF)
12        die("%s: end of file\n", filename);
13    }
14    return val;
15  }

 1  static unsigned int get_number255(FILE *fp, unsigned int maxval) {
 2    unsigned int val = get_number(fp);
 3    return (255*val+maxval/2)/maxval;
 4  }

 1  static void read_image(void) {
 2    ...
 3    case '2':
 4      /* Plain PGM */
 5      maxval = get_number(fp);
 6      for (i = 0; i < logo_height; i++)
 7        for (j = 0; j < logo_width; j++)
 8          logo_data[i][j].red = logo_data[i][j].green =
 9		     logo_data[i][j].blue = get_number255(fp, maxval);
10      break;
11      ...
12  }
\end{lstlisting}
    \vspace{-3mm}
\end{figure*}

\subsection{Examples of False Negatives}

In Listing~\ref{fig:fn1}, the function \verb|parse_encap_seg6| defines an array with 256 elements and passes it to \verb|netlink_parse_rtattr_nested|, along with the parameter \verb|max| set to 256. 
This function then forwards both arguments to \verb|netlink_parse_rtattr|, which eventually calls \lstinline|memset(tb, 0, sizeof(struct rtattr *) * (max + 1))|, resulting in an out-of-bounds write. 
This bug is successfully detected by \toolname{} but missed by both BugBot and CodeRabbit. 
In practice, the functions involved in this case span multiple files, each containing thousands of lines of code. Due to limited context length, both BugBot and CodeRabbit fail to recover the inter-procedural dependencies required to detect this vulnerability.

In Listing~\ref{fig:fn2}, the function \verb|get_number| sets \verb|val| to 0 if \verb|fgetc| reads the character \verb|'0'| from the input file. This value is then returned and propagated to the function \verb|read_image|, where it is assigned to the variable \verb|maxval|. Subsequently, \verb|maxval| is passed as the second argument to \verb|get_number255|, where it is used as a divisor without any prior validation. While this bug is correctly identified by \toolname{}, both BugBot and CodeRabbit fail to detect it. Detecting this issue requires interprocedural data flow tracking between the variables \verb|val| in \verb|get_number| and \verb|maxval| in \verb|get_number255|, along with reasoning about the value range of \verb|val| under different input conditions. Both BugBot and CodeRabbit struggle with such complex, context-dependent vulnerability scenarios.

\subsection{Examples of False Positives}

In Listing~\ref{fig:fp-bugbot}, the variable \verb|buf| is assigned the return value of the function \verb|pfile->cb.translate_include|. The function then calculates the length of the content in \verb|buf| using \verb|strlen(buf)|, and sets the next position to \verb|'\n'|. BugBot reports a buffer overflow for this operation. However, \verb|strlen(buf)| only describes the length of valid characters in \verb|buf| up to the null terminator, and does not necessarily reflect the actual allocated size of the buffer. Without enough context regarding the allocation of \verb|buf|, we cannot conclusively determine that \verb|strlen(buf)| represents the actual length of the buffer.

\begin{figure*}[t]
\vspace{-3mm}
\begin{lstlisting}[
        caption={The OOB false positive example reported by Cursor BugBot.},
        label={fig:fp-bugbot}
    ]
 1  bool _cpp_stack_file (cpp_reader *pfile, _cpp_file *file, include_type type, location_t loc) {
 2    char *buf = nullptr;
 3    ...
 4    if (!file->header_unit && type < IT_HEADER_HWM 
 5        && type != IT_INCLUDE_NEXT 
 6        && pfile->cb.translate_include)
 7    buf = (pfile->cb.translate_include(pfile, pfile->line_table, loc, file->path));
 8    ...
 9    size_t len = strlen (buf);
10    buf[len] = '\n';
11    cpp_buffer *buffer = 
12      cpp_push_buffer (pfile, reinterpret_cast<unsigned char *> (buf), len, true);
13    buffer->to_free = buffer->buf;
14    ...
15  }
\end{lstlisting}
\vspace{-4mm}
\end{figure*}

\begin{figure*}[t]
\vspace{-2mm}
\begin{lstlisting}[
        caption={The OOB false positive example reported by CodeRabbit.},
        label={fig:fp-coderabbit}
    ]
 1  FileNamesTable* UTIL_createExpandedFNT(const char* const* inputNames, size_t nbIfns, int followLinks) {
 2    unsigned nbFiles;
 3    char* buf = (char*)malloc(LIST_SIZE_INCREASE);
 4    char* bufend = buf + LIST_SIZE_INCREASE;
 5    ...
 6    size_t ifnNb, pos;
 7    for (ifnNb = 0, pos = 0; ifnNb < nbFiles; ifnNb++) {
 8      fileNamesTable[ifnNb] = buf + pos;
 9      if (buf + pos > bufend) { free(buf); free((void*)fileNamesTable); return NULL; }
10      pos += strlen(fileNamesTable[ifnNb]) + 1;
11    }
12    ...
13  }    
 
\end{lstlisting}
\vspace{-3mm}
\end{figure*}

In Listing~\ref{fig:fp-coderabbit}, the variable \verb|buf| is allocated a fixed memory region of size \verb|LIST_SIZE_INCREASE|, and \verb|bufend| is set to point to the end of this buffer. In the loop, each element of \verb|fileNamesTable| is assigned an offset within \verb|buf| using \verb|fileNamesTable[ifnNb] = buf + pos|. Before accessing the memory, the program checks \verb|if (buf + pos > bufend)| to ensure the access stays within bounds. CodeRabbit flags this comparison as a potential undefined behavior due to possible pointer overflow in \verb|buf + pos|. However, \verb|pos| is incremented gradually based on the length of the input filenames using \verb|pos += strlen(fileNamesTable[ifnNb]) + 1|. Triggering an overflow in \verb|buf + pos| would require \verb|pos| to approach the maximum value of \verb|size_t|, which in practice would require an unrealistically large number of large input strings (e.g., more than $10^{18}$). Therefore, this report is a false positive, as such extreme conditions are virtually impossible to reach in realistic scenarios.

\end{document}